\documentclass[useAMS,usenatbib]{mn2e}

\input psfig.sty
\usepackage{xspace}
\usepackage{graphicx}
\usepackage{amsmath,amssymb}

\newcommand {\apj} {ApJ}
\newcommand {\apjl} {ApJL}

\newcommand {\mnras} {MNRAS}

\newcommand {\aap} {A\&A}

\newcommand {\nat} {Nature}
\newcommand {\araa} {ARA\&A}

\newcommand {\pasp} {PASP}

\newcommand {\etal} {et~al.~}
\def \spose#1{\hbox  to 0pt{#1\hss}}  
\newcommand {\lta} {\mathrel{\spose{\lower 3pt\hbox{$\sim$}}\raise  2.0pt\hbox{$<$}}}
\newcommand {\gta} {\mathrel{\spose{\lower  3pt\hbox{$\sim$}}\raise 2.0pt\hbox{$>$}}}

\newcommand {\ha}  {\ifmmode H\alpha \else H$\alpha $ \fi} 
 
\newcommand {\kms} {\ifmmode  \,\rm km\,s^{-1} \else $\,\rm km\,s^{-1}  $ \fi }
\newcommand {\kpc} {\ifmmode  {\rm kpc}  \else ${\rm  kpc}$ \fi  }  
\newcommand {\pc} {\ifmmode  {\rm pc}  \else ${\rm pc}$ \fi  }  
\newcommand {\Msun} {\ifmmode {\rm M_{\odot}} \else ${\rm M_{\odot}}$ \fi} 
\newcommand {\Zsun} {\ifmmode {\rm Z_{\odot}} \else ${\rm Z_{\odot}}$ \fi} 
\newcommand {\yr} {\ifmmode yr^{-1} \else $yr^{-1}$ \fi} 
\newcommand {\hMsun} {\ifmmode h^{-1}\,\rm M_{\odot} \else $h^{-1}\,\rm M_{\odot}$ \fi}

\def\q3{q_{3}}

\usepackage[usenames]{color}

\def\lick{UCO/Lick Observatory, Department of Astronomy and Astrophysics, 
  University of California, Santa Cruz, CA 95064, USA}
\def\ucsb{Dept. of Physics, University of California, 
  Santa Barbara, CA 93106, USA}
\def\kipac{Kavli Institute for Particle Astrophysics and Cosmology, 
 Stanford University, 452 Lomita Mall, Stanford, CA 94035, USA}
\def\utah{Department of Physics and Astronomy, University of Utah, 
  Salt Lake City, UT 84112, USA}
\def\kapteyn{Kapteyn Astronomical Institute, University of Groningen, 
  P.O.Box 800, 9700 AV Groningen, The Netherlands}
\def\oxford{Department of Physics, University of Oxford, 
  Keble Road, Oxford, OX1 3RH, UK}
\def\cambridge{Institute of Astronomy, University of Cambridge,
  Madingley Rd, Cambridge, CB3 0HA, UK}
\def\mpia{Max Planck Institute for Astronomy, K\"onigstuhl 17, 69117, Heidelberg, Germany}
\def\auckland{Department of Statistics, The University of Auckland, 
Private Bag 92019, Auckland 1142, New Zealand}

\def\duttonemail{\tt dutton@mpia.de}
\def\packard{Packard Research Fellow}

\title [The IMF of massive bulges] {The SWELLS survey
  - V.  A Salpeter stellar initial mass function in the bulges of
  massive spiral galaxies}

\author[Dutton \etal]{%
  Aaron~A.~Dutton$^{1,2}$\thanks{\duttonemail}
  Tommaso~Treu$^2$\thanks{\packard}, 
  Brendon~J.~Brewer$^{2,3}$, 
  Philip~J.~Marshall$^{4}$,
\newauthor{%
  M.~W.~Auger$^{2,5}$,
  Matteo Barnab\`e$^{6}$,
  David~C.~Koo$^{7}$, 
  Adam~S.~Bolton$^{8}$, }
\newauthor{%
  Leon~V.~E.~Koopmans$^{9}$ }\\
  $^1$\mpia\\
  $^2$\ucsb\\
  $^3$\auckland\\
  $^4$\oxford\\
  $^5$\cambridge\\
  $^6$\kipac\\
  $^7$\lick\\ 
  $^8$\utah\\
  $^9$\kapteyn
}

\begin{document}
             
\date{Accepted 2012 October 22.  Received 2012 October 12; in original form 2012 June 19}
             
\pagerange{\pageref{firstpage}--\pageref{lastpage}}\pubyear{2012}

\maketitle         

\label{firstpage}


\begin{abstract}
  Recent work has suggested that the stellar initial mass function
  (IMF) is not universal, but rather is correlated with galaxy stellar
  mass, stellar velocity dispersion, or morphological type. In this
  paper, we investigate variations of the IMF within individual
  galaxies. For this purpose, we use strong lensing and gas kinematics
  to measure independently the normalisation of the IMF of the bulge
  and disk components of a sample of 5 massive spiral galaxies with
  substantial bulge components taken from the SWELLS survey. We find
  that the stellar mass of the bulges are tightly constrained by the
  lensing and kinematic data. A comparison with masses based on
  stellar population synthesis models fitted to optical and near
  infrared photometry favors a Salpeter-like normalisation of the IMF.
  Conversely, the disk masses are less well constrained due to
  degeneracies with the dark matter halo, but are consistent with
  Milky Way type IMFs in agreement with previous studies.  The disks
  are submaximal at 2.2 disk scale lengths, but due to the
  contribution of the bulges, the galaxies are baryon dominated at
  2.2 disk scale lengths.
  Globally, our inferred IMF normalisation is consistent with that
  found for early-type galaxies of comparable stellar mass
  ($>10^{11}$M$_\odot$). Our results suggest a non-universal IMF
  within the different components of spiral galaxies, adding to the
  well-known differences in stellar populations between disks and
  bulges.
\end{abstract}


\setcounter{footnote}{1}


\section{Introduction}
\label{sec:intro}

The stellar initial mass function (IMF) is a fundamental quantity in
many areas of astrophysics. From a theoretical standpoint,
understanding the origin of the IMF from first principles is essential
to develop a complete theory of star formation
\citep[e.g.,][]{McKee-Ostriker2007}.  From a phenomenological
standpoint, the IMF is a defining property of any stellar population,
essential for computing quantities such as stellar mass from
observables, and for characterising their evolutionary history.

Traditionally, the main source of empirical evidence regarding the
stellar IMF has been our own Milky Way, where individual stars can be
identified and counted, first by \citet{Salpeter1955}.  In the decades
since then, only relatively small variations of the IMF have been
found within our own Milky Way, despite enormous variations in the
physical conditions within star-forming regions. These results have
been generally interpreted as evidence that the IMF is more or less
universal---i.e. it is insensitive to properties of the gas and dust
in which stars form \citep[e.g.,][]{Bastian2010}. If indeed the
IMF is universal, the kinematics of spiral \citep{Bell-deJong2001} and
elliptical galaxies \citep{Cappellari2006}, as well as lensing data
\citep[][hereafter SWELLS-III]{Brewer2012} rule out IMFs like
Salpeter's implying relatively ``heavy'' mass-to-light ratios, in
favour of ``lighter'' IMFs like that measured by \citet{Kroupa2001}
and \citet{Chabrier2003}.

However, recently a number of independent extragalactic studies have
found significant deviations from the IMF as measured in the Milky
Way. The observations are based on a variety of independent
techniques, ranging from gravitational lensing (\citealt{Treu2010},
\citealt{Auger2010imf}, \citealt{Spiniello2011},
\citealt{Brewer2012}), to stellar kinematics of elliptical galaxies
\citep[e.g.,][]{Dutton-Mendel-Simard2012,
  Dutton-Maccio-Mendel-Simard2012, Cappellari2012}, and to spectral
diagnostics of stellar populations \citep{vanDokkum-Conroy2010,
  vanDokkum-Conroy2011,Spiniello2012, Conroy-vanDokkum2012}.

Based on these observations, it appears that the IMF may depend on the
stellar mass of the galaxy and hence on the cosmological time at which
the stars formed, possibly reflecting the evolving physical conditions
in the expanding universe. Or perhaps it could depend on the stellar
velocity dispersion, reflecting the depth of the potential well. It is
also possible that the IMF might depend on the morphological type of
the galaxy, in the sense that early-type galaxies might have
``heavier'' IMFs than their spiral counterparts. However, the
dependency of the distribution of morphological types on stellar mass
\citep[e.g.,][]{Blanton-Moustakas2009}, as well as the presence of two
very distinct stellar populations in the bulge and disks of typical
galaxies \citep[e.g.,][]{Wyse1997} make it unclear whether the more
important parameter is the overall stellar mass, or stellar velocity
dispersion, or the morphological type.

In this paper we aim to gather some insight into the physical origin
of the non-universality of the IMF by looking for variations {\it
  within} individual spiral galaxies. Is the normalisation of the IMF
the same for bulge and disk, just varying with total stellar mass? Or
is there one universal normalisation for disk-like stellar populations
(presumably Chabrier-like) and a ``heavier'' (Salpeter-like) one for
the older and more metal rich stellar populations found in massive
bulges and spheroids? Clearly the two hypotheses are not mutually
exclusive, as a combination of both could be at play.

Our strategy to constrain the IMF normalisation consists of comparing
stellar masses derived from dynamics and strong lensing to those
derived from stellar population synthesis (SPS) models.  The main
challenge of this approach -- and of all other dynamical approaches --
is that lensing and dynamics are sensitive to the total mass, and
therefore one needs to disentangle the stellar mass from the
non-baryonic dark matter. For disk-dominated galaxies it is well known
that rotation curves can be fitted with a wide range of stellar
mass-to-light ratios, from zero to maximum disk
\citep[e.g.,][]{vanAlbada-Sancisi1986, Dutton2005}, the so called
disk-halo degeneracy \citep[see however][]{Amorisco-Bertin2010}. For
bulge-dominated systems, the situation is similar in principle (the
so-called bulge-halo degeneracy). However, if the stellar mass of the
bulge dominates the inner parts of the gravitational potential, it is
possible to get an actual measurement (rather than an upper limit),
with only modest assumptions about the density profile of the dark
matter halo, e.g., inspired by the results of cosmological numerical
simulations (\citealt{Treu-Koopmans2002a, Koopmans-Treu2003, Treu2010,
  Auger2010imf, Cappellari2012}; see also \citealt{Bertin1994}, and
references therein).

In order to study the IMF of the bulge and disk component we apply the
lensing and dynamical technique to the sample of five massive
($V_{2.2}\sim 250 - 300 \kms$, $M_{\rm SPS}^{\rm Chab}\sim
10^{11}\Msun$) spiral lens galaxies in the SWELLS Survey
\citep[][hereafter SWELLS-I]{Treu2011} with significant bulges
(bulge-to-disk ratio $\gta 0.5$ assuming a universal IMF) and for
which ionised gas kinematic data are available. SWELLS is a dedicated
survey to the study of lensing spiral disk galaxies. Its main
properties are summarized in Section~2. More details can be found in
the paper by \citet{Treu2011}.
Thus, our mass models are constrained by the projected mass within the
Einstein radius measured by strong gravitational lensing, by the
rotation curve of the outer disk $\sim 1-3$ scale lengths from ionised
gas kinematics, and the high resolution surface brightness profiles
which are decomposed into a de Vaucouleurs bulge and an exponential
disk. We derive stellar masses of the bulge and disk for a variety of
assumptions about the structure of the dark matter halo, and compare
the results with the estimates from stellar population synthesis
models to infer the normalisation of the IMF independently for the two
components.

This paper is organised as follows. In \S2 we present the
observational constraints. In \S3 we present the mass models. In \S4
we present of our main results. In \S5 we discuss possible sources of
systematic errors. In \S6 we conclude with a brief summary.
Throughout, we assume a flat $\Lambda$CDM cosmology with present day
matter density, $\Omega_{\rm m}=0.3$, and Hubble parameter, $H_0=70
\rm\,km\,s^{-1}\,Mpc^{-1}$.

\begin{figure*} 
\includegraphics[width=0.98\linewidth]{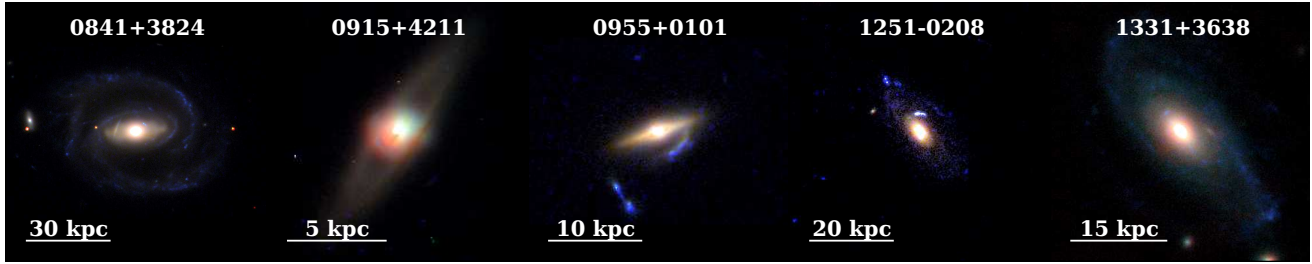}
\caption{Postage stamp images of the galaxies analysed in this
  paper. The images are obtained from multicolour HST images as
  described in SWELLS-I and III.}
\label{fig:montage}
\end{figure*}


\begin{table*}
\begin{center}
  \caption{Basic properties of galaxies analyzed in this
    paper. All errors correspond to 1$\sigma$. Col. 1 lists the lens
    ID; Col. 2 the redshift of the deflector; Col.3 gives the velocity
    dispersion of the Singular Isothermal Ellipsoid (SIE) lens model
    from paper III; Cols. 4 and 5 give half-light sizes for the bulge
    (circularized) and disk (major axis) from papers I and III;
    Cols. 6, 7 and 8 give the stellar masses for the bulge, disk, and
    total, assuming a Chabrier (2003) IMF; Cols. 9 and 10 give the
    model (dust reddened) rest-frame $V$-band luminosities for the
    bulge and disk; Cols. 11 and 12 give the model (B-V) colors for
    the bulge and disk.}
\label{tab:basic}
\scriptsize
\begin{tabular}{lcccccccccccc}
  \hline
  \hline ID & $z_{\rm d}$ & $\sigma_{\rm SIE}$ & $R_{\rm bulge}$ & $R_{\rm disk}$ & $\log(M^{\rm Chab}_{\rm SPS,b})$ & $\log(M^{\rm Chab}_{\rm SPS,d})$ & $\log(M^{\rm Chab}_{\rm SPS,t})$ & $\log(L_{V,\rm b})$ & $\log(L_{V,\rm d})$ & $(B-V)_{\rm b}$ & $(B-V)_{\rm d}$ \\
                    &       & [km s$^{-1}$] & [kpc] & [kpc] & [$\Msun$] & [$\Msun$] & [$\Msun$] & [$L_{\odot,V}$] & [$L_{\odot,V}$] &\\
      (1)           & (2)   & (3)   & (4) & (5) & (6) & (7) & (8) & (9) & (10) & (11) & (12) \\
  \hline
  J0841+3824 & $0.116$ & $251.2\pm4.4$ & $2.42$ & $23.43$& $11.05\pm0.09$ & $11.23\pm0.09$ & $11.45\pm0.07$ & $10.54\pm0.04$ & $10.84\pm0.04$ & 0.79 & 0.70\\
  J0915+4211 & $0.078$ & $195.7\pm2.2$ & $1.56$ & $\phantom{1}4.15$ & $10.60\pm0.09$ & $10.17\pm0.09$ & $10.74\pm0.08$ & $10.03\pm0.05$ & $\phantom{1}9.73\pm0.05$ & 0.81 & 0.72\\
  J0955+0101 & $0.111$ & $238.4\pm7.3$ & $1.62$ & $\phantom{1}3.72$ & $10.63\pm0.09$ & $10.17\pm0.09$ & $10.76\pm0.07$ & $10.00\pm0.04$ & $\phantom{1}9.73\pm0.04$ & 0.87 & 0.73\\
  J1251$-$0208&$0.224$ & $203.0\pm2.6$ & $1.68$ & $12.01$& $10.68\pm0.07$ & $10.96\pm0.07$ & $11.14\pm0.06$ & $10.16\pm0.04$ & $10.68\pm0.03$ & 0.82 & 0.65\\
  J1331+3638 & $0.113$ & $248.1\pm4.4$ & $2.86$ & $12.46$& $10.89\pm0.10$ & $10.46\pm0.10$ & $11.03\pm0.07$ & $10.38\pm0.04$ & $10.34\pm0.04$ & 0.78 & 0.47\\

  \hline \hline
\end{tabular}
\medskip\\
\end{center}
\end{table*}

\section{Sample selection and data}
\label{sec:data}

In this section we describe the sample selection and give a brief
description of the data used in this paper. The reader is referred to
papers SWELLS-I, SWELLS-II \citep{Dutton2011-SW}, SWELLS-III, and
SWELLS-VI (Dutton et al., in preparation) for more details on the
SWELLS selection and data.

\subsection{Sample selection}

As detailed in paper I, the parent SWELLS sample is selected from the
Sloan Digital Sky Survey (SDSS) spectroscopic database based on the
detection of emission lines at multiple redshifts within the $3''$
fiber and on having photometric axis ratio $<0.6$. Multiband Hubble
Space Telescope imaging is used to confirm the lensing hypothesis. The
sample of twenty confirmed lenses spans a broad range of bulge to
total stellar mass ratio (from 0.1 to 0.9) for a fixed universal IMF,
and a range in stellar mass of over one decade ($\sim10^{10}$M$_\odot$
to $10^{11.5}$M$_\odot$; for a Chabrier IMF).

An important question is whether the SWELLS sample is representative
of the overall population of spiral galaxies within the same
ellipticity and stellar mass limits. Paper I shows that the
size-stellar mass relation of the SWELLS sample is indistinguishable
from that of a control sample of SDSS-selected galaxies using the same
criteria.  Paper VI will investigate the selection function in more
detail using kinematic data to construct the Tully-Fisher, Faber
Jackson and Fundamental Plane correlations.

The subsample analysed in this paper is selected from the SWELLS
survey to include all the massive spiral galaxies with significant
bulges for which gas-kinematic rotation curves are available. We
require the stellar mass in the disk to be less than two times the
stellar mass in the bulge, a criterion that is met by all but two of
the lenses in our sample (we also exclude one system that has been
shown to have a pseudo-bulge) although only 5 of these systems also
have gas rotation curves available. A montage of our selected
subsample of galaxies is shown in Figure~\ref{fig:montage}. The
sub-sample includes some (but not all) of the most massive SWELLS
lenses. Quantitatively, the subsample spans a range in lensing
velocity dispersion of 196-251 \kms, i.e. galaxies that are more
massive than our own Milky Way, and spanning the 230\kms threshold
below which paper III concluded that a global Salpeter IMF is on
average ruled out by the data. Basic properties of the subsample
analysed in this paper are given in Table~\ref{tab:basic}.

\subsection{Strong Lensing}
Strong lensing models for our sample galaxies are presented in
SWELLS-III. The parameter that is most directly constrained by strong
lensing is the circularised Einstein radius $b$ \citep[][and
  references therein]{Treu2010review}. This is defined to be the
radius inside which the average surface density is equal to the
critical density for strong lensing:
\begin{equation}
 \frac{M_{\rm proj}(b)}{\pi b^2} =
\Sigma_{\rm crit} \equiv \frac{c^2}{4\pi G} \frac{D_{\rm s}}{D_{\rm ds} D_{\rm d}}.
\end{equation}
Here $D_{\rm s}$ is the angular diameter distance from the observer to
the source, $D_{\rm d}$ is the angular diameter distance from the
observer to the deflector (i.e., the lens), and $D_{\rm ds}$ is the
angular diameter distance from the deflector to the source.  The
critical density thus depends only on the distances to the lens and
source, which are known for all our lens systems from the SDSS
redshifts and our adopted cosmology.

For comparison to kinematics it is convenient to express the lensing
results in terms of circular velocity or velocity dispersion, where
$V_{\rm SIE}=\sqrt{2}\sigma_{\rm SIE}$. However, since the mass
profiles are in general not isothermal, the conversion from projected
mass into circular velocity at the Einstein radius is non trivial. For
a total mass profile steeper than isothermal, which is the case for
our galaxies, the circular velocity at the Einstein radius is larger
than the nominal circular velocity of the singular isothermal
ellipsoid (SIE) model.

Thus in our mass models in this paper we fit directly for $b$, rather
than the mass projected within $b$ (due to covariance between $M_{\rm
  proj}(b)$ and $b$), or the derived circular velocity/velocity
dispersion at the Einstein radius.

\subsection{Gas kinematics}

We obtained major axis long-slit spectra for all five galaxies with
the Low Resolution Imaging Spectrograph \citep[LRIS;][]{Oke1995} on the
Keck I 10-m telescope between November 2008 and April 2011.  Typical
exposure times were 60 minutes, with seeing conditions of FWHM $\sim
1.0$ arcsec.  On the red side we used the 600/7500 line grating which
gives a pixel scale of $1.26$ \AA~px$^{-1}$. With a 1\arcsec width
slit the resulting spectral resolution is $\simeq 4.2$ \AA,
corresponding to velocity dispersion resolution of $\sim 70\kms$ at
the wavelength of H$\alpha$.  We adopt spatial samplings of $\simeq
1\farcs1 - 1\farcs5$ (5-7 pixels) corresponding to $\sim 1$ data point per
seeing FWHM.

We measured rotation curves by fitting Gaussians to one-dimensional
spectra extracted along the slit. To improve centroiding accuracy we
fitted the redshift of neighbouring emission lines simultaneously. The
primary set of emission lines are H$\alpha$ 6563, [NII] 6583, 6548.
The line flux ratios between the two [NII] lines were fixed at the
expected values. The velocity dispersions of the line pairs were also
imposed to be the same, but were allowed to be different than
H$\alpha$. The continuum was fitted with a 2nd order polynomial.

The observed rotation velocities are lowered from the true circular
velocities by several effects. Firstly, there is a $\sin(i)$ term due
to the inclination, $i$, of the galaxy with respect to the line of
sight.  Secondly, there is the (potential) position angle offset of
the slit from the kinematic major axis. Thirdly there is beam smearing
due to finite slit width and seeing.  These effects are taken into
account in our modelling. We create model velocity fields and extract
flux weighted rotation velocities inside a 1\arcsec width slit, which
has a position angle offset from the major axis. For this calculation
we assume the line emitting gas traces the stellar disk. This
assumption is the major source of systematic uncertainty in our
models.  To minimise the impact of the uncertainties in the beam
smearing model we exclude from our fits the inner few arcseconds of
the kinematic data. We also exclude data points where the rotation
curve is strongly asymmetric (such as due to the bar in J0841+3824),
or where the signal to noise is low (such as in inter spiral arm
regions).
We treat the disk inclination angle and slit position angle offsets as
nuisance parameters, allowing them to vary over suitably chosen small
ranges.  We determine an initial guess for the inclination from the
disk minor-to-major axis ratio $q_{\rm d}$ assuming $\cos(i) = \sqrt{(
  q_{\rm d}^2 - q_0^2)/(1-q_0^2)}$, where $q_0\simeq 0.2$
\citep[e.g.,][]{Hall2012} is the intrinsic disk thickness.

\subsection{Bulge and disk structural parameters}
The light profiles are decomposed into bulges and disks as described
in papers SWELLS-I and III. The bulge light profile is assumed to
follow a S\'ersic $n=4$ (de Vaucouleurs) profile, while the disk
profile is assumed to be S\'ersic $n=1$ (exponential) profile. We note
that 0841+3824 is a barred spiral galaxy. However, the bar is
significantly fainter than the bulge ($\approx 2$ magnitudes) and
therefore it can be considered negligible from a dynamical standpoint,
within our desired level of precision, thus simplifying significantly
the analysis.

\subsection{Stellar Population Synthesis Masses}
The photometry (typically in 4 bands, BVIK) is used to compute SPS
masses using the method of \citet{Auger2009} together with
\citet{Bruzual-Charlot2003} SPS models. The SPS masses for the bulge
and disk are computed using both \citet{Chabrier2003} and
\citet{Salpeter1955} IMFs.  We consider stellar populations described
by 5 parameters: the total stellar mass~$M_{\rm SPS}$, the population
age~$A$, the exponential star formation burst timescale $\tau$, the
metallicity~$Z$ and the reddening due to dust, $\tau_V$. We employ a
uniform prior requiring $9 \leq {\rm log_{10}} (M_{\rm SPS}/\Msun)
\leq 13$, the age is constrained such that star formation began at
some (uniformly likely) time between $1 \leq z \leq 5$, $\tau$ has an
exponential prior with characteristic scale 1 Gyr, and we impose
uniform priors on the logarithms of the metallicity and dust
extinction such that $-4 \leq {\rm log_{10}} Z \leq -1.3$ and $-2 \leq
{\rm log_{10}} \tau_V \leq 0.3$. We note that the priors are the same
for the bulge and the disk components but are sufficiently
conservative that they do not bias our results. The posterior PDF is
sampled as described in \citet{Auger2009}.  The stellar mass estimates
are the mean and standard deviation of the marginalised posterior PDF.

As discussed by several authors, especially for the old and mostly
dust free populations found in massive bulges, the uncertainties in
stellar mass inferred from colors are dominated by the IMF
normalization \citep[see, e.g., recent discussion by][and references
therein]{Newman2012}.  The degeneracies between ages and metallicities
typical of color-based inferences mostly cancel out when inferring
stellar masses \citep{Bell-deJong2001,Auger2009}. Similarly, the
effects of using different stellar population synthesis models are
typically of order 0.05-0.1 dex for stellar populations not dominated
by thermally pulsating AGB stars \citep[e.g.,][]{Treu2010}.

\section{Reference Mass model}

In this section we define our reference model, which we will use in
Section~\ref{sec:res} to interpret the data. This is a very general
model and should be regarded as the one providing the definitive
results of this paper. However, in order to test the robustness of our
conclusions, in Section~\ref{sec:systematics} we define and apply to
our data a full battery of alternative models, including ones with
adiabatic contraction and expansion, as well as models characterised
by dark matter cores, by different implementations of the
cosmologically motivated dark matter halos introduced here, and models
without dark matter. As we will show, our conclusions are robust with
respect to the choice of model. Therefore, the reader pressed for time
can skip Section~\ref{sec:systematics} and jump directly to the
Summary after Section~\ref{sec:res}.

\subsection{Overview}
Our reference model is one with two baryonic components, a spherical
bulge and a thin disk, and a generic dark matter halo, that includes
as a subset standard profiles motivated by cosmological
simulations. We refer to this model as the ``free'' dark matter model.

We model the bulge with a \citet{Hernquist1990} profile, which is
parametrised by its mass, $M_{\rm bulge}$, and a half-mass radius,
$r_{\rm bulge}$.  We model the disk with an exponential profile, which
is parametrised by its mass, $M_{\rm disk}$, and a half-mass radius
$r_{\rm disk}$.  We model the dark matter halo with a generalised
spherical \citet[][NFW]{NFW1997} profile:
\begin{equation}
\rho(r) \propto (r/r_{\rm s})^{-\gamma} (1 + r/r_{\rm s})^{-3 +\gamma},
\end{equation}
where $\gamma$ is the inner slope (NFW corresponds to $\gamma=1$), and
$r_{\rm s}$ is the scale radius.  The normalisation is determined by
the virial velocity, $V_{200}$, which is defined at a radius,
$R_{200}$, enclosing a mean density of 200 times the critical density
of the Universe at redshift zero. The relation between virial
velocity, virial radius, and virial mass is thus given by
\begin{equation}
  \frac{V_{200}}{[\kms]}=\frac{R_{200}}{[h^{-1} \rm kpc]}=\left(G\frac{M_{200}}{[h^{-1}\Msun]}\right)^{1/3},
\end{equation}
where $h=H_0/100 \,{\rm \kms Mpc^{-1}}$ and $G\approx 4.302\times
10^{-6}\, {\rm km^2 s^{-2}\, kpc\, \Msun^{-1}}$. 
This reference model thus has 7 parameters (4 for the baryons, 3 for
the dark matter).

\begin{figure*}
\centering
\includegraphics[width=0.48\linewidth]{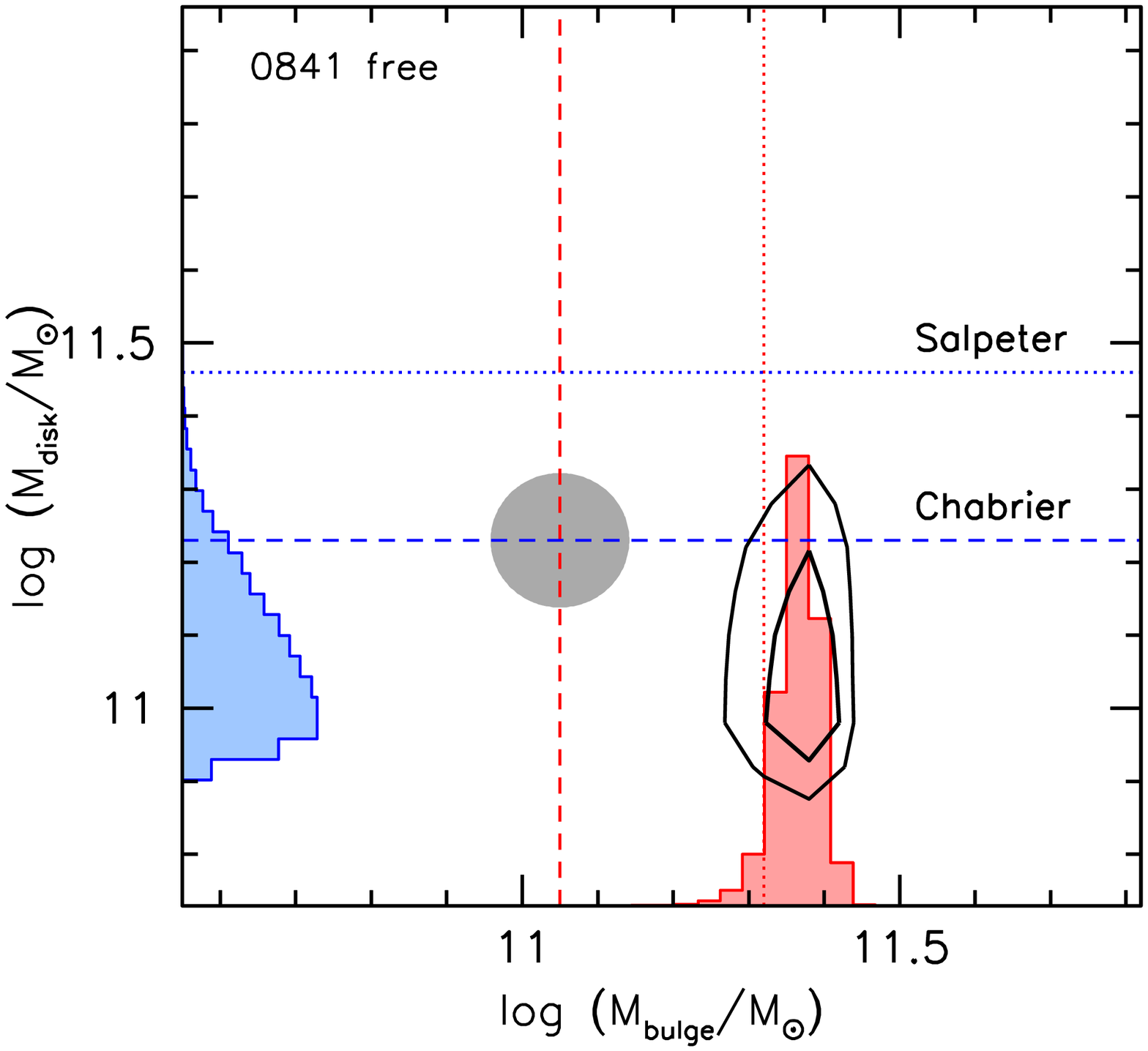}
\includegraphics[width=0.48\linewidth]{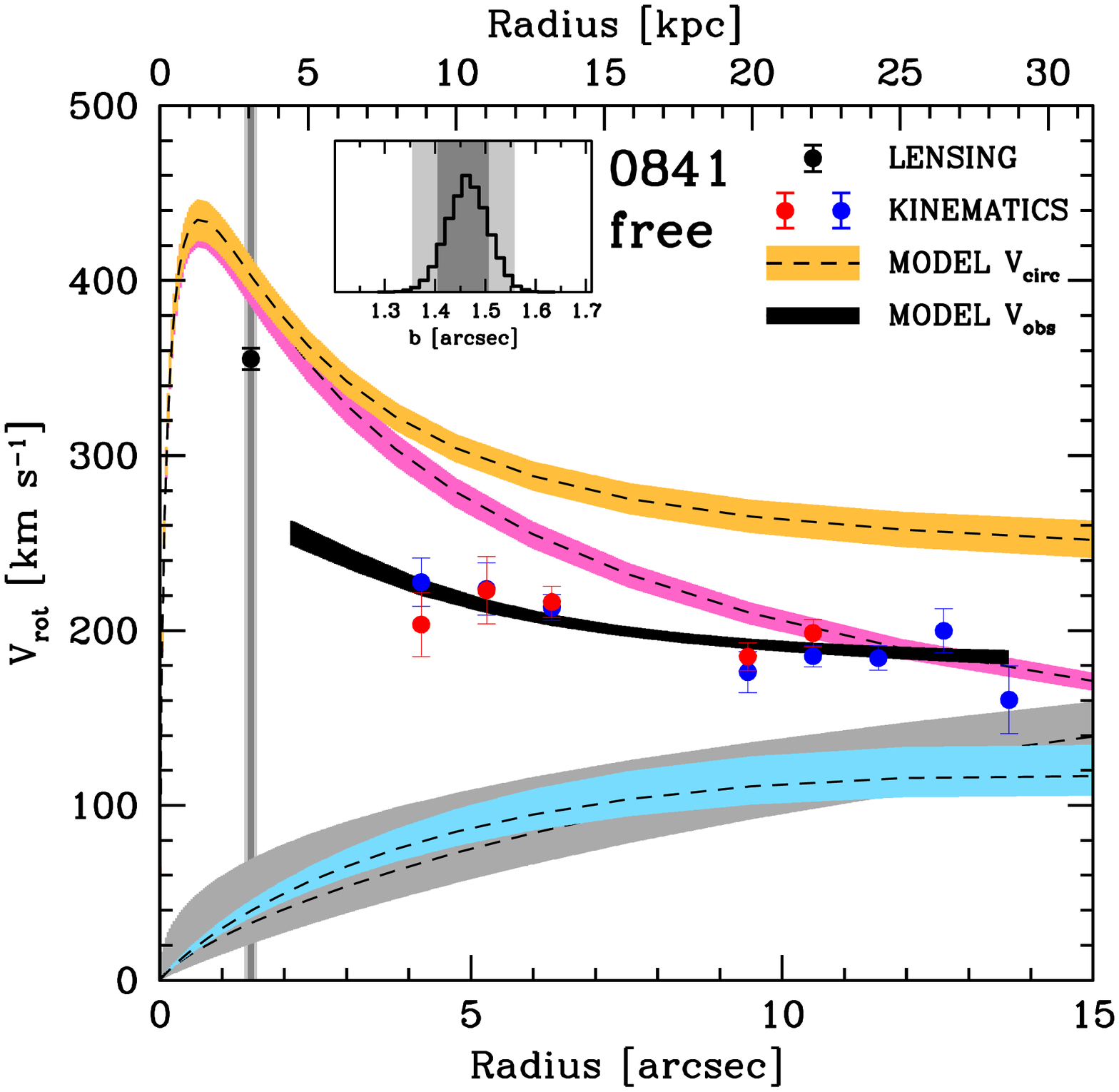}
\includegraphics[width=0.48\linewidth]{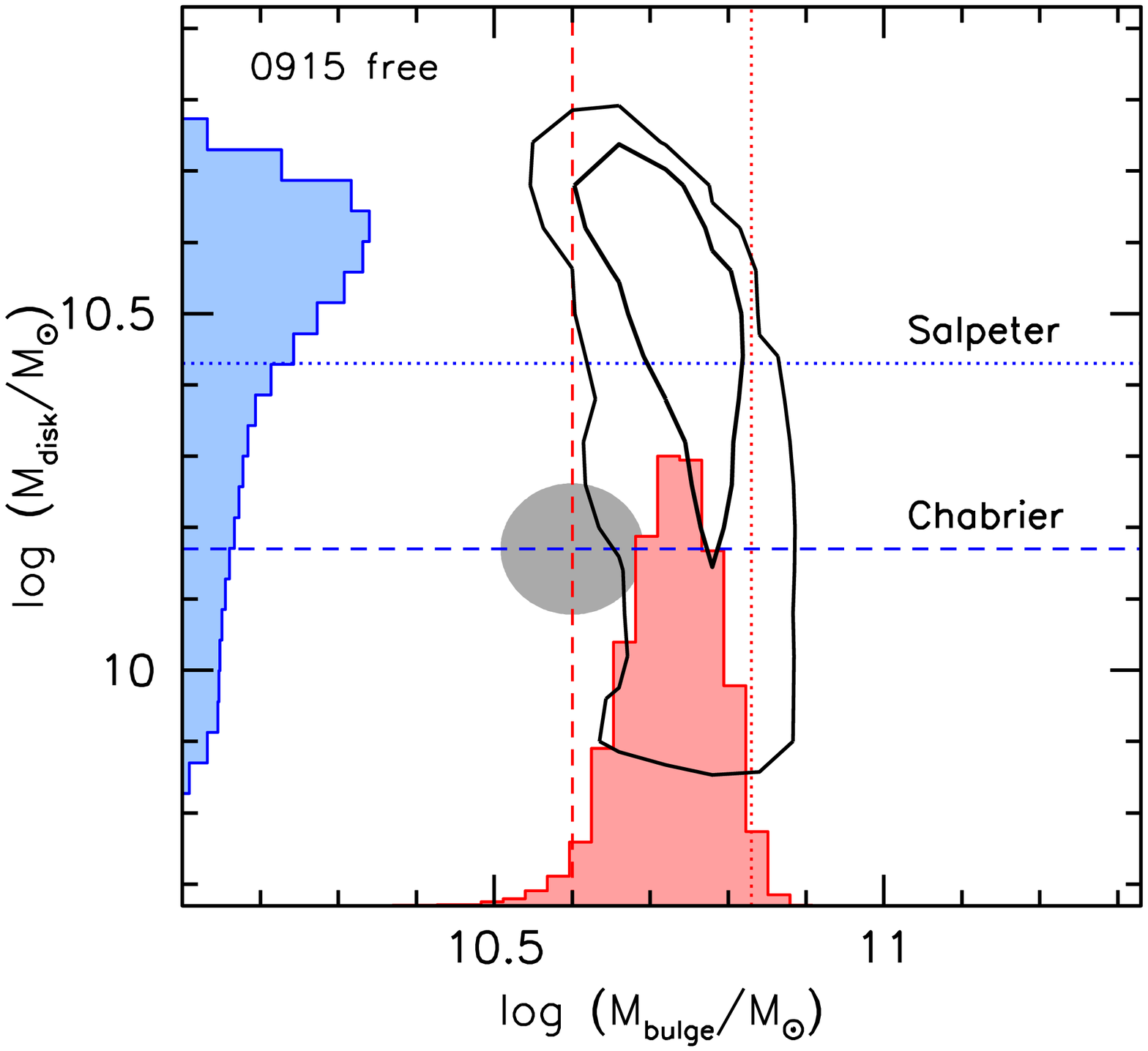}
\includegraphics[width=0.48\linewidth]{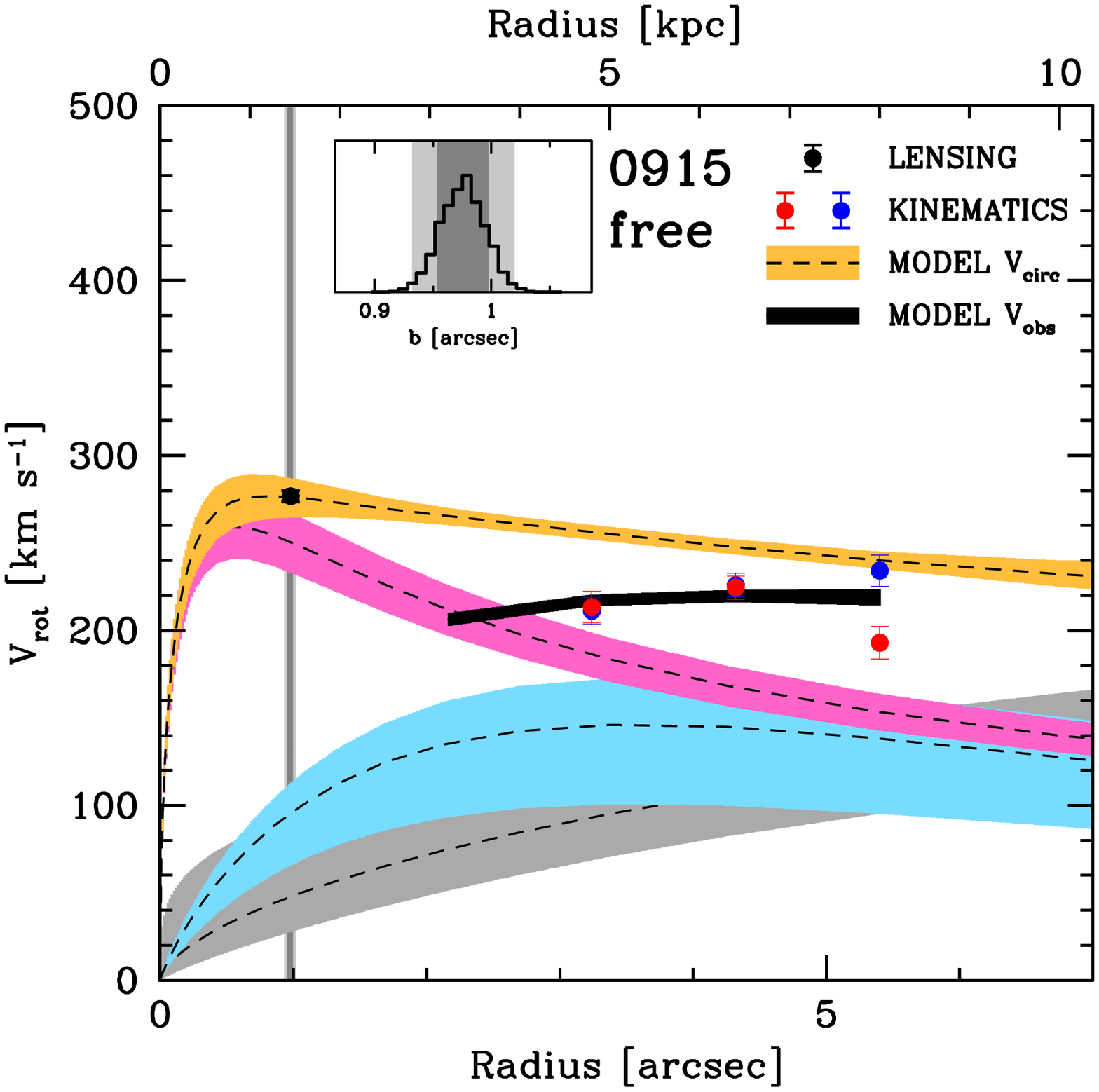}
\caption{Mass model results for a free dark matter halo. {\it Left:}
  Joint constraints on bulge and disk masses. The contours enclose
  68\% and 95\% of the probability. The red and blue histograms show
  the marginalised probability distributions for the bulge and disk
  masses, respectively. The dashed and dotted lines correspond to the
  median SPS masses assuming Chabrier and Salpeter IMFs,
  respectively. For reference, the grey ellipse shows the 1$\sigma$
  uncertainties on these SPS masses assuming a universal Chabrier IMF.
  {\it Right:} Data model comparison. The black shaded region shows
  the 68\% confidence region of the ``observed'' model (including
  inclination and beam smearing effects). This should be compared to
  the red and blue points, which correspond to the observed rotation
  velocities from the receding and approaching sides of the galaxy,
  respectively. In the inset panels histograms show the distribution
  of model Einstein radii, while the shaded bands correspond to the
  $1$ and $2\sigma$ constraints on the observed Einstein radii. These
  grey bands are repeated in the main panel. For reference the black
  point shows the circular velocity of the SIE model (note: this is
  not used in the fit). In the main panels, the orange shaded region
  shows the 68\% constraint on the total circular velocity
  profile. The magenta, cyan, and grey shaded regions show the
  circular velocity profiles due to the bulge, disk, and dark matter
  halo, respectively. The bulge masses are well constrained due to the
  declining nature of the circular velocity curves, while the disk
  masses are less well constrained, due to degeneracies with the dark
  matter halo.}
\label{fig:vm}
\end{figure*}

\begin{figure*}
\centering
\includegraphics[width=0.44\linewidth]{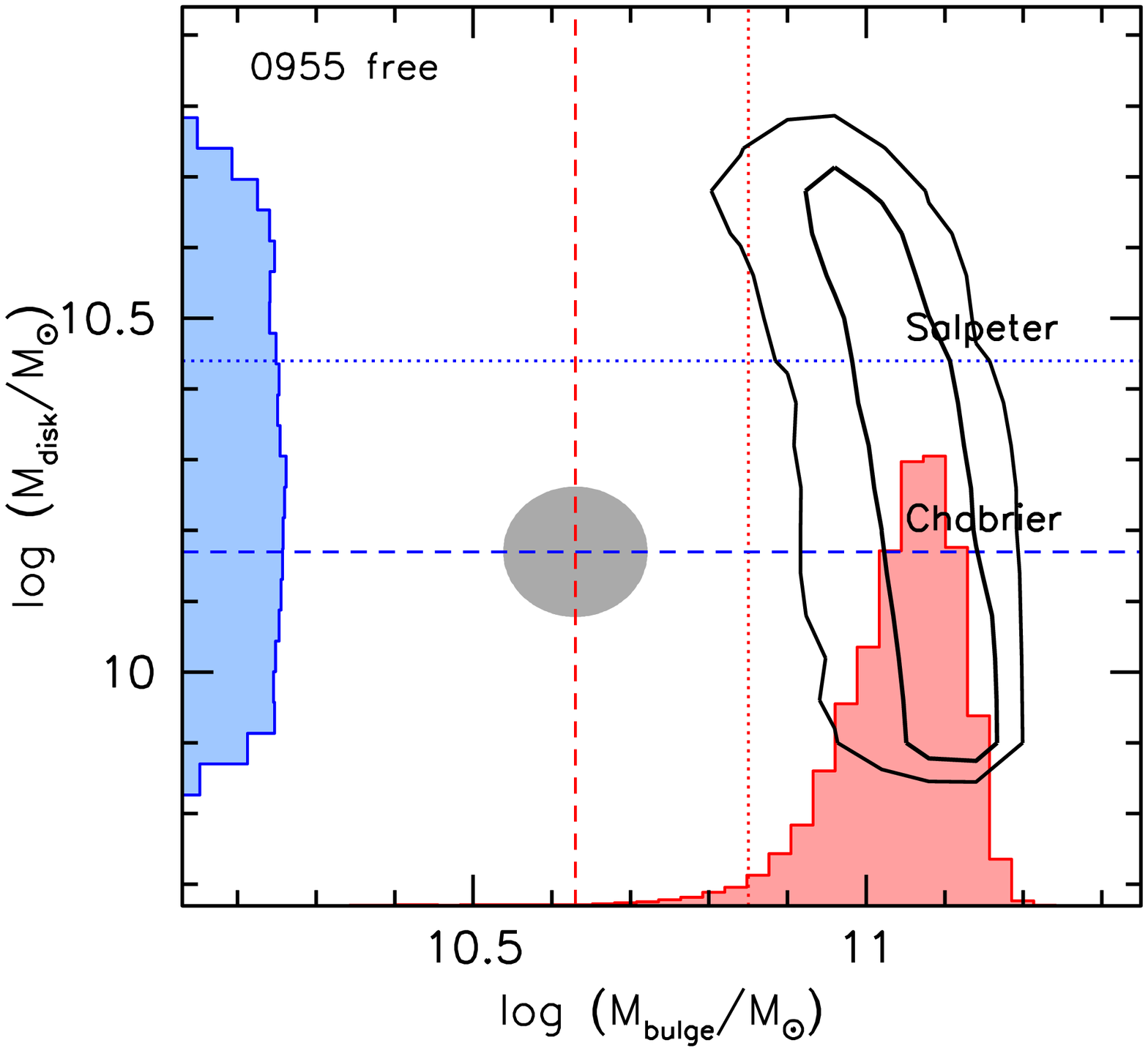}
\includegraphics[width=0.44\linewidth]{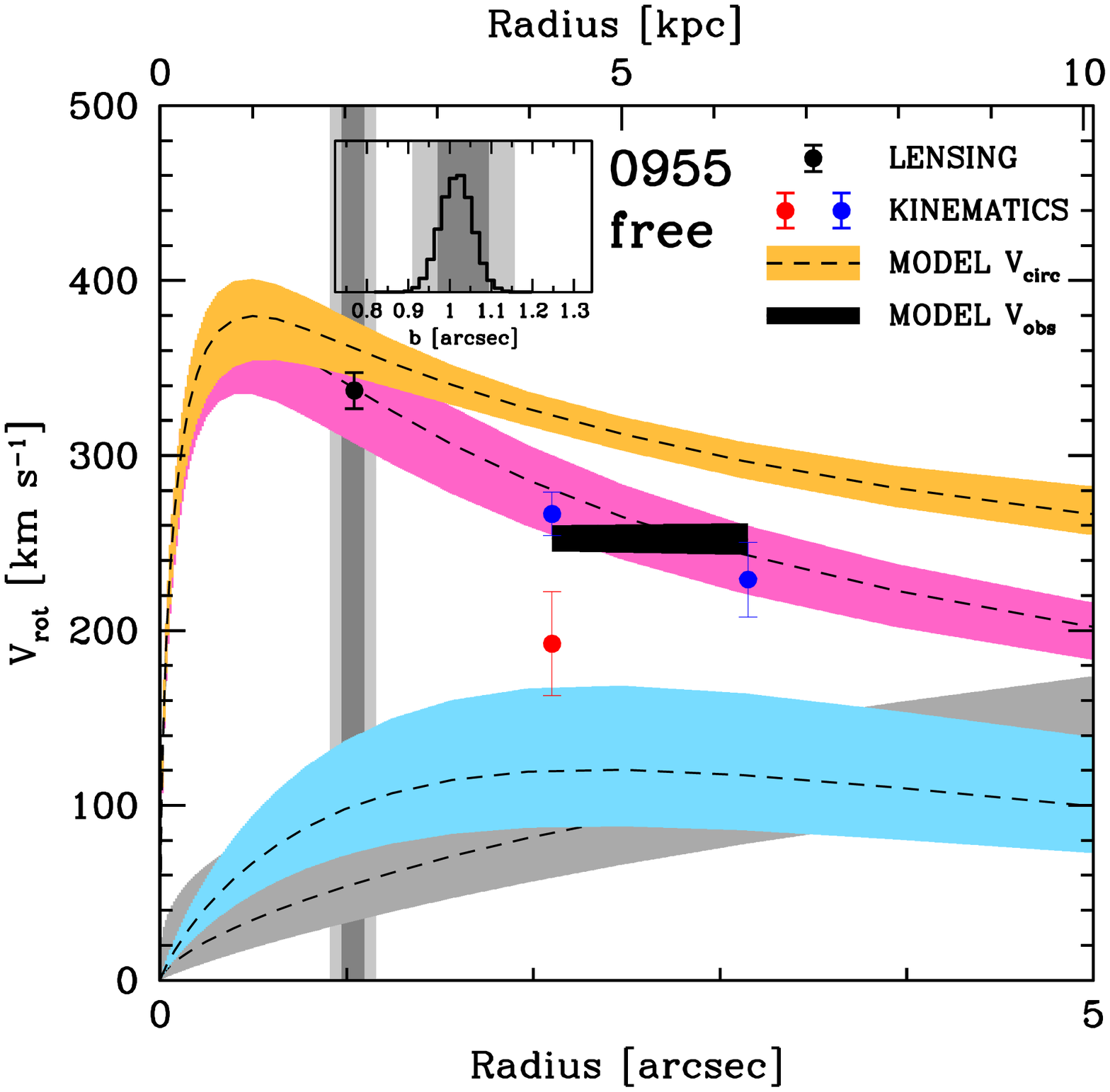}
\includegraphics[width=0.44\linewidth]{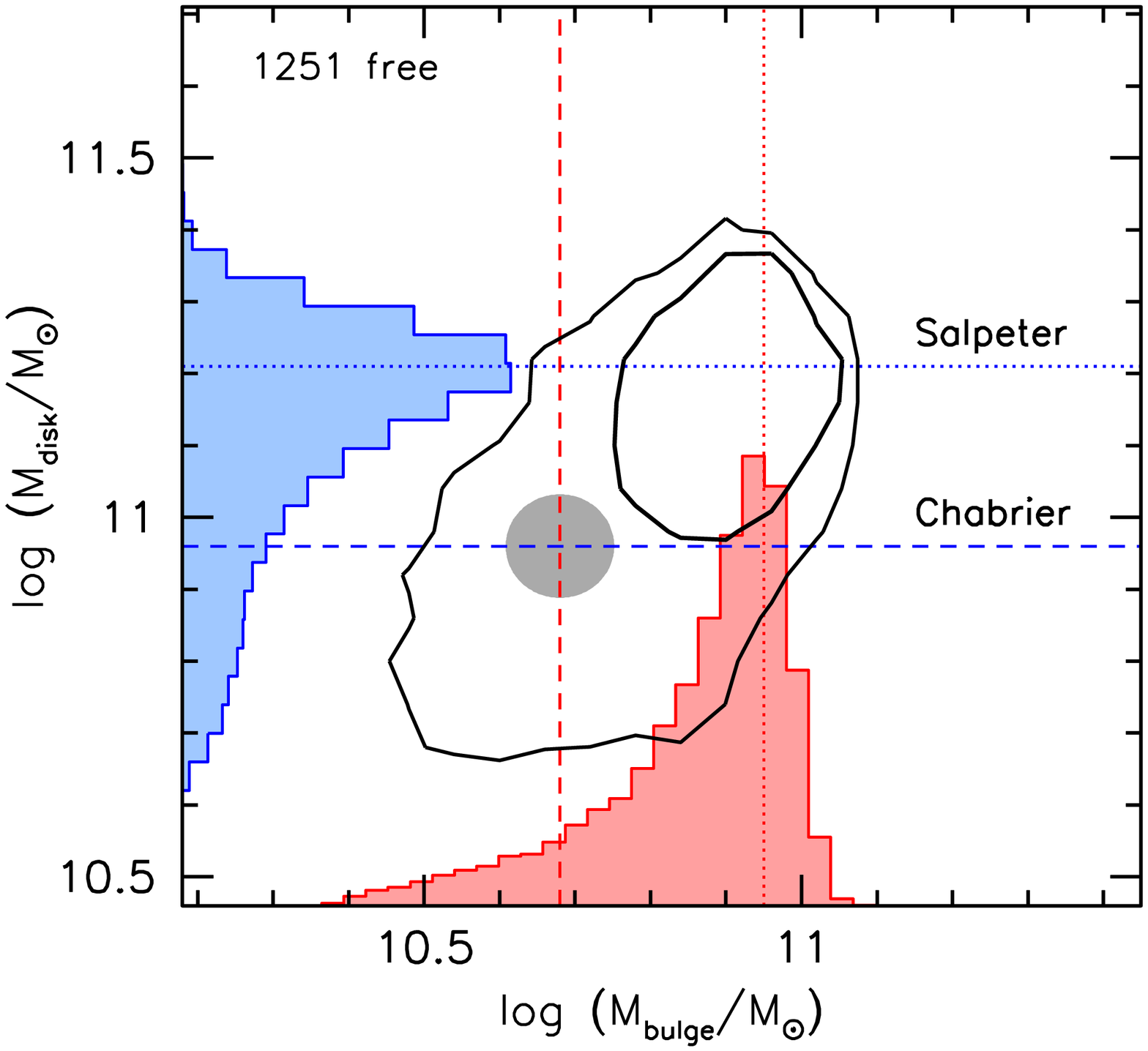}
\includegraphics[width=0.44\linewidth]{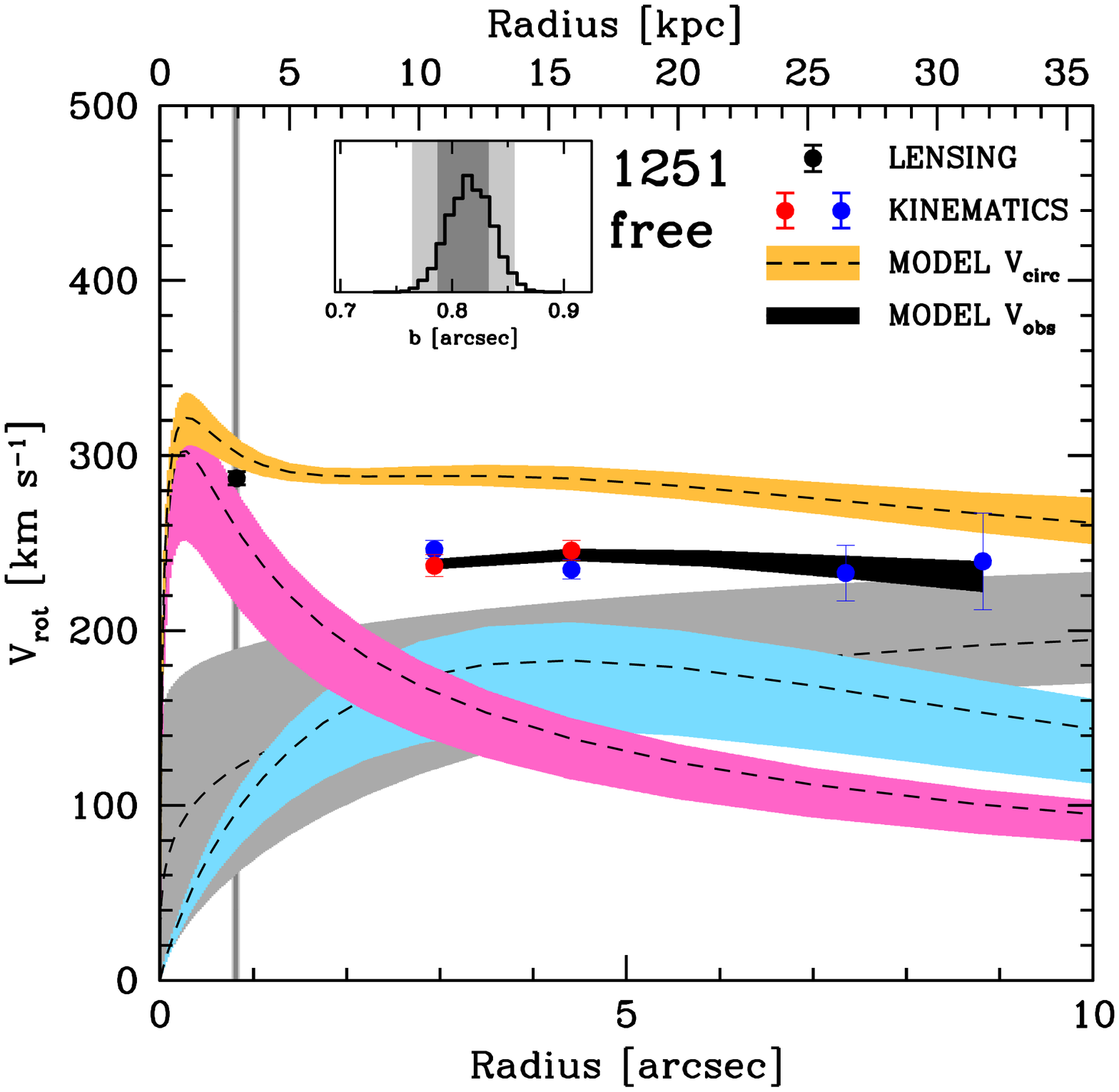}
\includegraphics[width=0.44\linewidth]{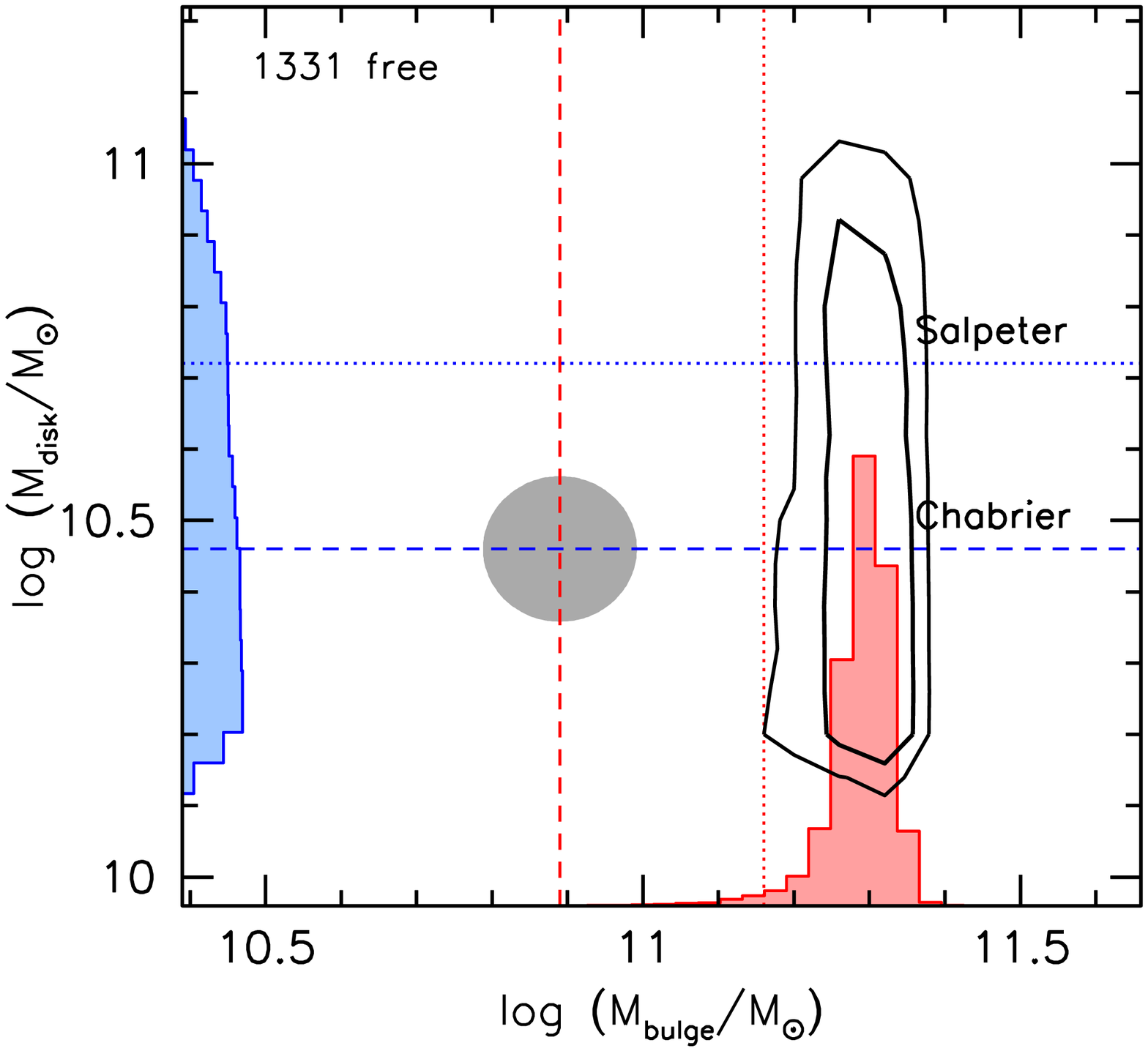}
\includegraphics[width=0.44\linewidth]{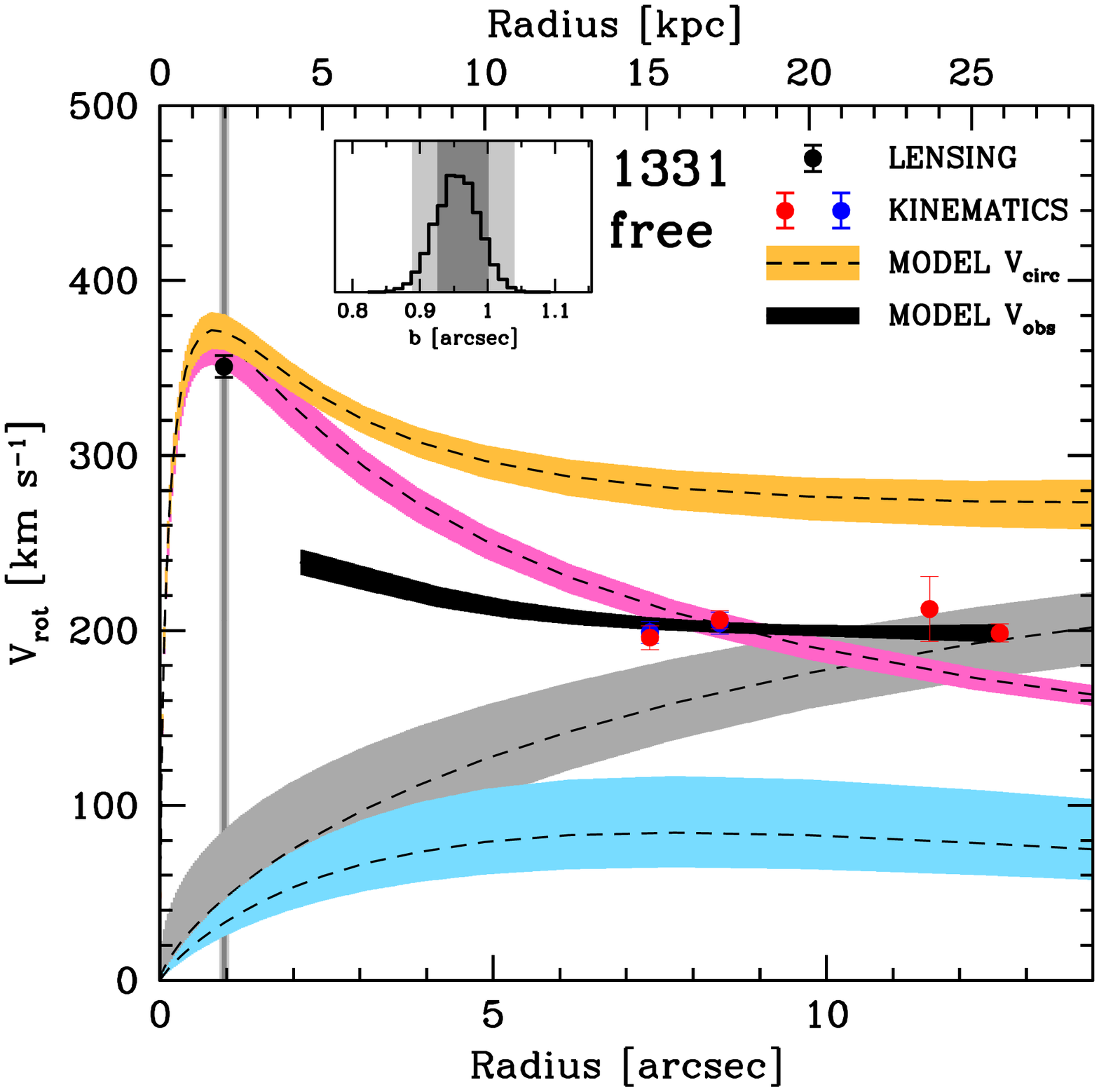}
\addtocounter{figure}{-1}
\caption{Continued}
\label{fig:vm2}
\end{figure*}

\subsection{Constraints and priors} 
\subsubsection{Baryons} 

The disk and bulge sizes are fixed to the values obtained from our
photometric bulge-disk decompositions (from SWELLS I and III). This
assumes negligible mass-to-light ratio gradients in the bulge and disk
and leaves 2 free parameters, the masses of the bulge and the disk.
For these we adopt uniform priors in $\log_{10}(M)$ over an interval
bracketing the range of plausible values as inferred from SPS
masses. In practice, we adopt a lower limit of half the SPS mass
assuming a Chabrier IMF, and an upper limit of twice the SPS mass
assuming a Salpeter IMF. For old stellar populations, this upper limit
corresponds to a power-law IMF with slope $\simeq -3$.

\subsubsection{Dark matter} 

For the inner slope of the dark matter density profile we adopt a
uniform prior in the interval $0 \le \gamma < 2$. The lower limit
corresponds to a cored halo, while the upper limit corresponds to
isothermal, which mimics strong halo contraction. 

For the scale radius of the halo we adopt a prior based on the
concentration mass relation of $\Lambda$CDM haloes from
\citet{Maccio2008}. The median relation is given by
\begin{equation}
\log_{10}(c) =
   0.830-0.098\log_{10}\left(\frac{M_{200}}{10^{12}h^{-1}\Msun}\right).
\end{equation}
The concentration, $c$, is defined to be the ratio between the virial
radius and the radius where the slope of the density profile is $-2$:
$c = R_{200}/r_{-2}$.  For an NFW halo $r_{-2}=r_{\rm s}$, but for a
generalised NFW halo $r_{-2}=r_{\rm s}/(2-\gamma)$.  We adopt a
Gaussian prior on $\log_{10}(c|M_{200})$ with standard deviation of
0.11 dex, which is the scatter in halo concentrations found in
cosmological simulations \citep{Maccio2008}. This choice of prior is
not critical to our results.  Its main purpose is to introduce more
freedom in the dark matter model than obtained by assuming a fixed
scale radius \citep[e.g.,][]{Treu2010,Cappellari2012}, and to ensure
the scale radii are not unphysically large or small.

For the virial velocity of the dark matter halo we adopt a uniform
prior in the interval: $V^{\rm min}_{200} \le V_{200} \le V^{\rm
  max}_{200} \kms$.  The lower limit to the virial velocity is
obtained by assuming the baryon fraction (inside the virial radius) is
equal to the cosmic baryon fraction, $f_{\rm bar}=0.16$, thus
\begin{equation}
V^{\rm min}_{200} = \left(\frac{GM_{\rm total}}{f_{\rm bar}/(1-f_{\rm
    bar})}\right)^{1/3} = \left(\frac{116.4}{\kms}\right)
\left(\frac{M_{\rm total}}{10^{11}\Msun} \right)^{1/3},
\end{equation}
where the total stellar mass $M_{\rm total}=M_{\rm bulge}+M_{\rm
  disk}$.  Similarly, the upper limit $V^{\rm max}_{200}$, corresponds
to a baryon-to-stars conversion efficiency of 1\%. This conservative
limit is motivated by satellite kinematics, weak lensing and halo
abundance matching, which find conversion efficiencies of $\gta 2.5\%$
in massive galaxies \citep{Dutton2010}.

\subsection{Mass model fitting} 
The model, parametrized by $\theta$, is fitted to the kinematic and
lensing data simultaneously using a Bayesian Markov Chain Monte Carlo
approach similar to that used in previous SWELLS papers.  Specifically
we fit for the lensing Einstein radius, $b$, and the observed gas
rotation curve $V_{\rm rot}(r_i)$ using the standard likelihood
function:
\begin{equation}
\mathcal{L}(\theta)= \exp\left\{-\frac{ [b - b^{\rm mod}(\theta)] ^2}{2\sigma^2_{b}} 
-\sum_i\frac{ [V_{\rm rot}(r_i) - V_{\rm rot}^{\rm mod}(\theta)] ^2}{2\sigma^2_{V,i}} \right\}.
\end{equation}
As described above, our model for the gas rotation curve takes into
account beam smearing due to seeing and finite slit width.

\begin{table*}
\begin{center}
  \caption{Summary of stellar masses from fits to lensing and dynamics
    data with a free dark matter halo, and a comparison with SPS based
    masses. Cols 2,3, and 4 give the stellar masses (median together
    with 68\% posterior probability) of the bulge, disk and total
    (bulge+disk); Cols 5,6, and 7 give the IMF mismatch parameter
    $\alpha\equiv M_{\rm LD}/M^{\rm Chab}_{\rm SPS}$ (mean and
    standard deviation); Col 8 gives an upper limit to $\alpha$ within
    the Einstein radius using results directly from SWELLS-III.}
\label{tab:mass}
\begin{tabular}{lccccccc}
  \hline
  \hline Name & $\log(M_{\rm LD,bulge})$ & $\log(M_{\rm LD,disk})$ & $\log(M_{\rm LD,total})$ & $\log(\alpha_{\rm bulge})$ & $\log(\alpha_{\rm disk})$ & $\log(\alpha_{\rm total})$ & $\log(\alpha_{\rm lens})$\\
   (1) & (2) & (3) & (4) & (5) & (6) & (7) & (8)\\
  \hline 
  J0841+3824 & $11.37^{+0.02}_{-0.03}$ & $11.06^{+0.12}_{-0.09}$ & $11.54^{+0.05}_{-0.03}$ & $0.31\pm0.10$ & $-0.15\pm0.13$ & $0.10\pm0.08$ & $0.29\pm0.10$ \\
  J0915+4211 & $10.73^{+0.06}_{-0.06}$ & $10.50^{+0.14}_{-0.32}$ & $10.94^{+0.03}_{-0.07}$ & $0.13\pm0.11$ & $\phantom{-}0.27\pm0.24$ & $0.18\pm0.09$ & $0.09\pm0.10$ \\
  J0955+0101 & $11.06^{+0.06}_{-0.08}$ & $10.29^{+0.29}_{-0.27}$ & $11.14^{+0.03}_{-0.05}$ & $0.41\pm0.12$ & $\phantom{-}0.12\pm0.26$ & $0.37\pm0.09$ & $0.41\pm0.08$ \\
  J1251$-$0208& $10.90^{+0.07}_{-0.16}$ & $11.16^{+0.10}_{-0.21}$& $11.35^{+0.07}_{-0.18}$ & $0.18\pm0.15$ & $\phantom{-}0.15\pm0.17$ & $0.17\pm0.14$ & $0.33\pm0.07$ \\
  J1331+3638 & $11.29^{+0.03}_{-0.04}$ & $10.50^{+0.28}_{-0.23}$ & $11.37^{+0.05}_{-0.04}$ & $0.40\pm0.11$ & $\phantom{-}0.06\pm0.25$ & $0.34\pm0.09$ & $0.42\pm0.09$ \\
  Ensemble   &                       &                       &                       & $0.29\pm0.05$ & $-0.01\pm0.12$ & $0.24\pm0.04$ & \\ 
  \hline \hline
\end{tabular}
\medskip\\
\end{center}
\end{table*}

\section{Results} 
\label{sec:res}

The results presented in this section are based on models with the
``free'' dark matter halo described in Section 3. As discussed in
\S~\ref{sec:systematics}, none of our main results are sensitive to
the functional form of the dark matter halo that we adopt.

The results of our fits to the rotation curves and lensing Einstein
radii are shown in the right panels of Fig.~\ref{fig:vm}.  The black
shaded region shows samples of acceptable models, which should be
compared to the red and blue data points (which correspond to the
receding and approaching sides of the rotation curve, respectively).
The inset panels show the 1 and $2\sigma$ constraints on the Einstein
radius from strong lensing (SWELLS-III). The model Einstein radii are
shown as histograms.  The orange shaded regions show the 68\%
confidence regions on the total circular velocity. The magenta, blue
and grey shaded regions show the decomposition into bulge, disk, and
dark matter, respectively.  Overall the models fit the data remarkably
well and allow us to answer some specific questions.

\begin{figure}
\centering
\includegraphics[width=0.98\linewidth]{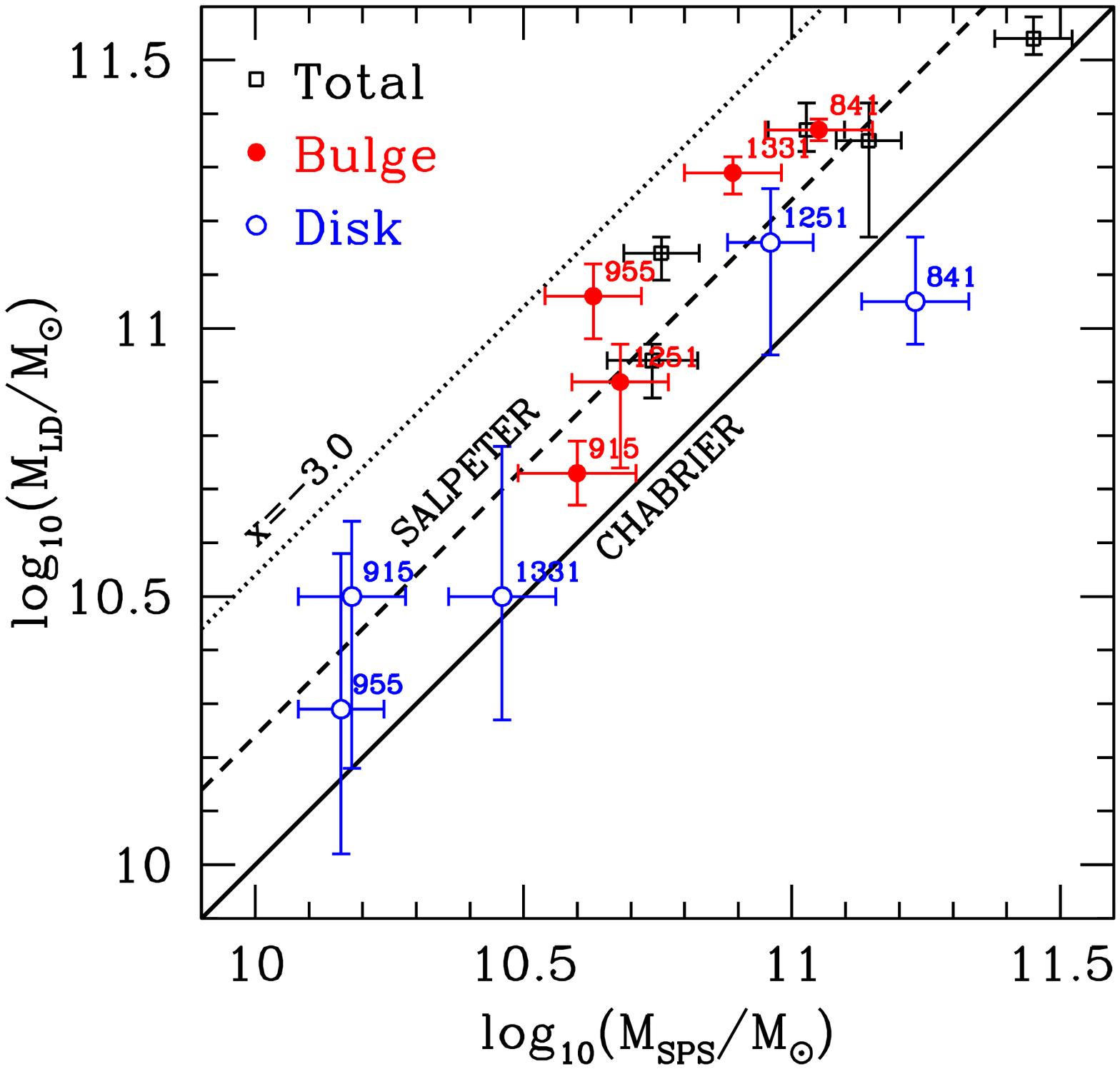}
\caption{Comparison between bulge (red, filled circles), disk (blue,
  open circles) and total (black, squares) stellar masses from lensing
  plus dynamics, $M_{\rm LD}$, with those from stellar population
  synthesis models assuming a Chabrier IMF, $M_{\rm SPS}$. The error
  bars enclose 68\% of the posterior probability.  The disk masses of
  our five galaxies are consistent with a Chabrier IMF, but the bulge
  (and total) masses favour an IMF a factor of $\simeq 2$ heavier than
  Chabrier.}
\label{fig:mm}
\end{figure}

\subsection{Is the IMF universal within individual galaxies?}

The posterior probability distribution functions (PDFs) for the bulge
and disk masses are shown in the left panels of Fig.~\ref{fig:vm} next
to the corresponding rotation curves.  The marginalised constraints on
the bulge, disk, and total stellar masses are given in
Table~\ref{tab:mass} (for reference, the corresponding masses from SPS
models are given in Table~\ref{tab:basic}).  In the reference ``free''
models, the disk stellar masses are somewhat loosely constrained due
to the well-known disk-halo degeneracy
\citep[e.g.,][]{vanAlbada-Sancisi1986}. However, the bulge masses are
well constrained because the inner part of the mass density profile is
too steep to be described by the dark matter halo.

The grey circles in Fig.~\ref{fig:vm} correspond to a universal
Chabrier IMF.  Three of the five galaxies in our sample are
inconsistent (at greater than $2\sigma$ level) with a universal
Chabrier IMF in {\it both} the bulge and disk.  A heavier IMF is
required for at least the bulge. Since Chabrier IMF seems to be in
general preferred for starforming galaxy disks, our data indicate that
the IMF may be non-universal {\it within galaxies}.

A comparison between the bulge and disk masses from our lensing plus
dynamics analysis, $M_{\rm LD}$, with those from SPS models, $M_{\rm
  SPS}$, is shown in Fig.~\ref{fig:mm}. For the disks, the LD masses
are consistent with a Chabrier IMF, in agreement with previous studies
of galactic disks \citep[e.g.,][]{Bell-deJong2001, Bershady2011,
  Dutton2011-IMF, Barnabe2012}.  However for the bulges, the LD masses
are a factor of $\simeq 2$ higher than predicted by a Chabrier IMF and
consistent with a Salpeter-like normalisation.  This latter result is
in agreement with studies of massive early-type galaxies
\citep[e.g.,][]{Auger2010imf, vanDokkum-Conroy2010, Spiniello2012,
  Dutton-Maccio-Mendel-Simard2012, Conroy-vanDokkum2012}.  It is also
interesting to note that the surface densities of the SWELLS bulges
are comparable to the highest density ETGs in the local universe,
which also favour IMFs with Salpeter-type normalisation
\citep{Dutton-Mendel-Simard2012}.

\begin{figure*}
\centering
\includegraphics[width=0.48\linewidth]{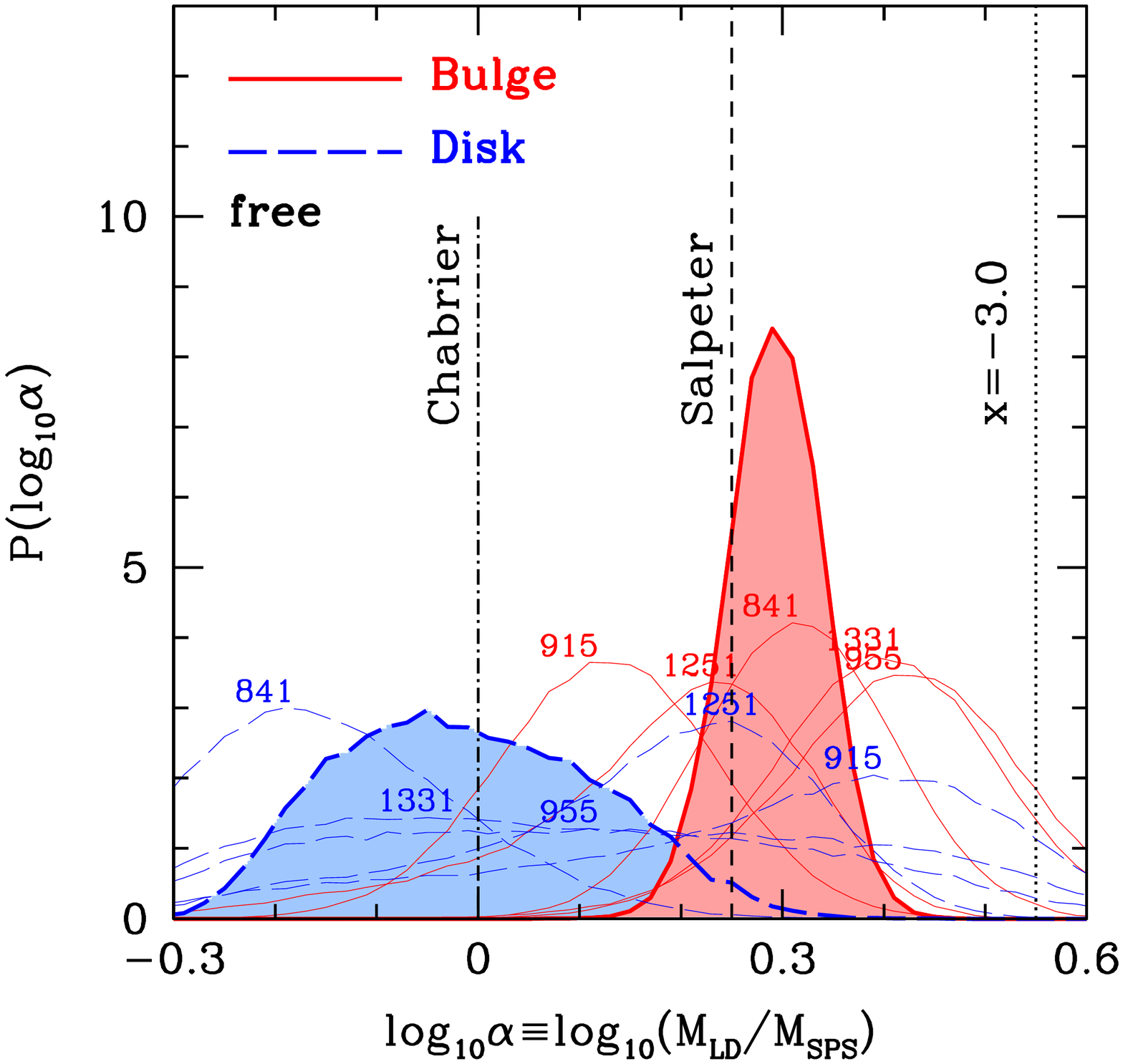}
\includegraphics[width=0.48\linewidth]{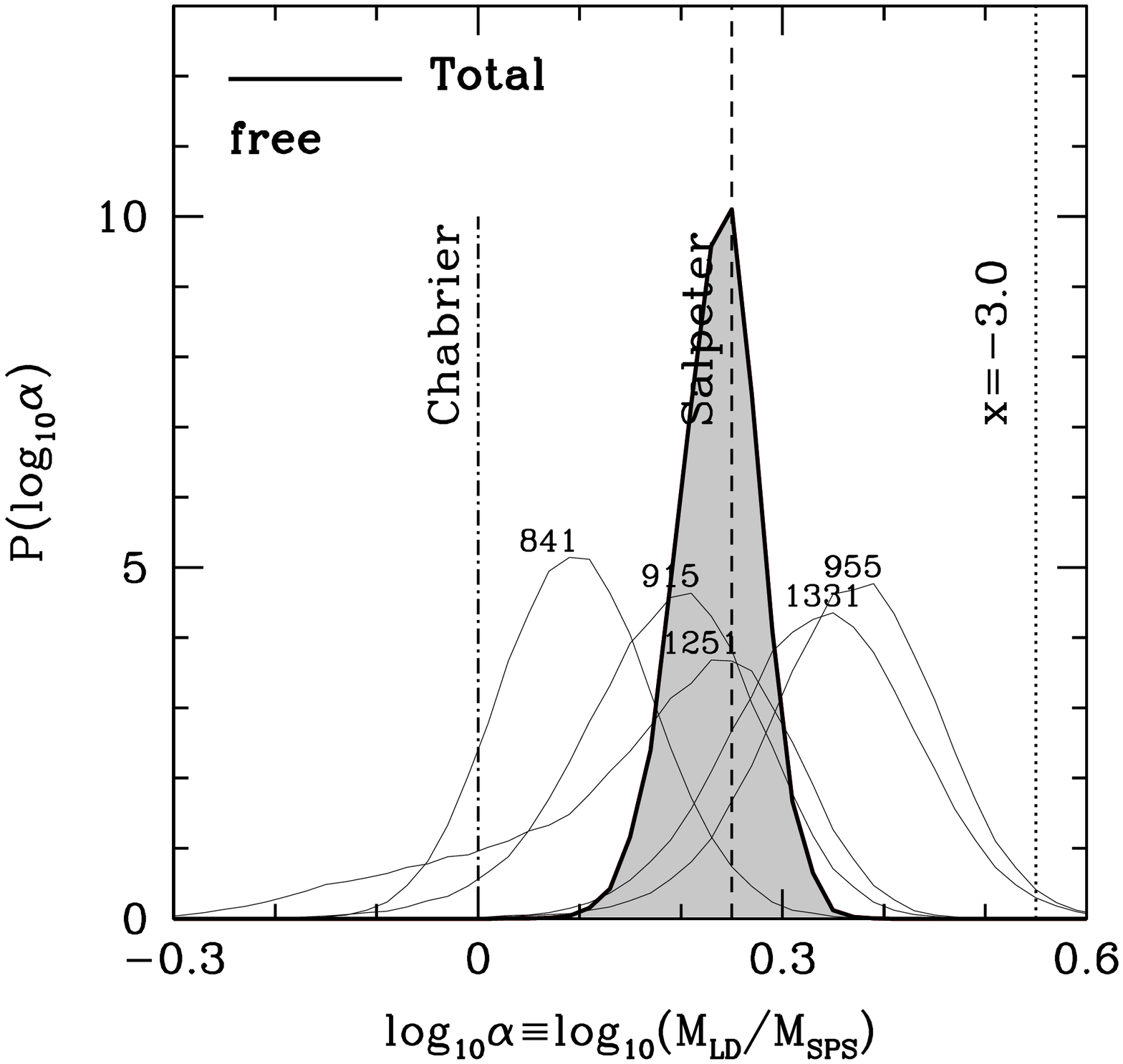}
\caption{Posterior distributions for the IMF mismatch parameter
  relative to a Chabrier IMF, $\alpha$. {\it Left:} for bulges (red,
  solid lines) and disks (blue, long dashed lines); {\it Right:} for
  total (bulge+disk) stellar mass. The bold histograms show the joint
  constraints on the mean $\alpha$. Disks are consistent with a
  Chabrier IMF, while bulges require an IMF roughly twice as heavy as
  a Chabrier IMF. The overall normalisation is close to Salpeter.}
\label{fig:alpha}
\end{figure*}

Fig.~\ref{fig:alpha} shows the PDFs for the ``IMF mismatch parameter''
\citet{Treu2010} $\alpha=M_{\rm LD}/M_{\rm SPS}$ quantifying the
relation between between LD masses and SPS masses.  Uncertainties on
both measurements are taken into account when deriving the posterior
distribution of $\alpha$. The mean and standard deviation of the PDFs
from Fig.~\ref{fig:alpha} are given in Table~\ref{tab:mass}.  On
average, the IMF of bulges is a factor of $\sim 2$ heavier than
Chabrier ($\log\alpha_{\rm bulge}=0.29\pm0.05$), while the IMF of the
disks is consistent with Chabrier ($\log\alpha_{\rm
  disk}=-0.01\pm0.12$).  We note that in 4 out of 5 galaxies the disk
masses are also consistent with a Salpeter IMF.  The total (bulge +
disk) masses have $\log\alpha_{\rm total}=0.24\pm0.04$, which is
consistent with a Salpeter-like normalisation of the IMF.

In SWELLS-III we concluded that a global IMF with $\alpha$ above the
Salpeter value is ruled out (at 98\% confidence) for galaxies with
lensing velocity dispersions below 230 km s$^{-1}$. This might seem at
odds with the conclusions of this paper. However, 3 out of the 5
galaxies studied here have velocity dispersion greater than 230 km
s$^{-1}$ (see Table~\ref{tab:basic}). The galaxies with higher
velocity dispersions all favor a bulge $\alpha$ values higher than the
Salpeter value, while the galaxies with lower velocity dispersions
favor bulge $\alpha$ values lower than the Salpeter value.

In addition, Table~\ref{tab:mass} gives the $\alpha$ parameter inside
the Einstein radius calculated using results presented in
SWELLS-III. Since these $\alpha$ are based on total masses, they are
upper limits. Thus for consistency they should be larger than the
$\alpha$ we derive in this paper. A direct comparison is complicated
because the $\alpha$'s are measured in different apertures, but
nevertheless, the $\alpha$ values for the bulges that we derive here
are fully consistent with the upper limits from SWELLS-III.

\subsection{Are disks sub-maximal?}
The rotation curves of spiral galaxies have contributions from the
stellar mass, gas mass, and dark matter. The relative contributions of
each are difficult to estimate from rotation curve modelling due to
well-known degeneracies. This led to the maximum disk hypothesis
\citep[e.g.,][]{vanAlbada-Sancisi1986} which proposes that the stellar
disk makes the maximum possible contribution to the rotation curve.

\begin{figure}
\centering
\includegraphics[width=0.98\linewidth]{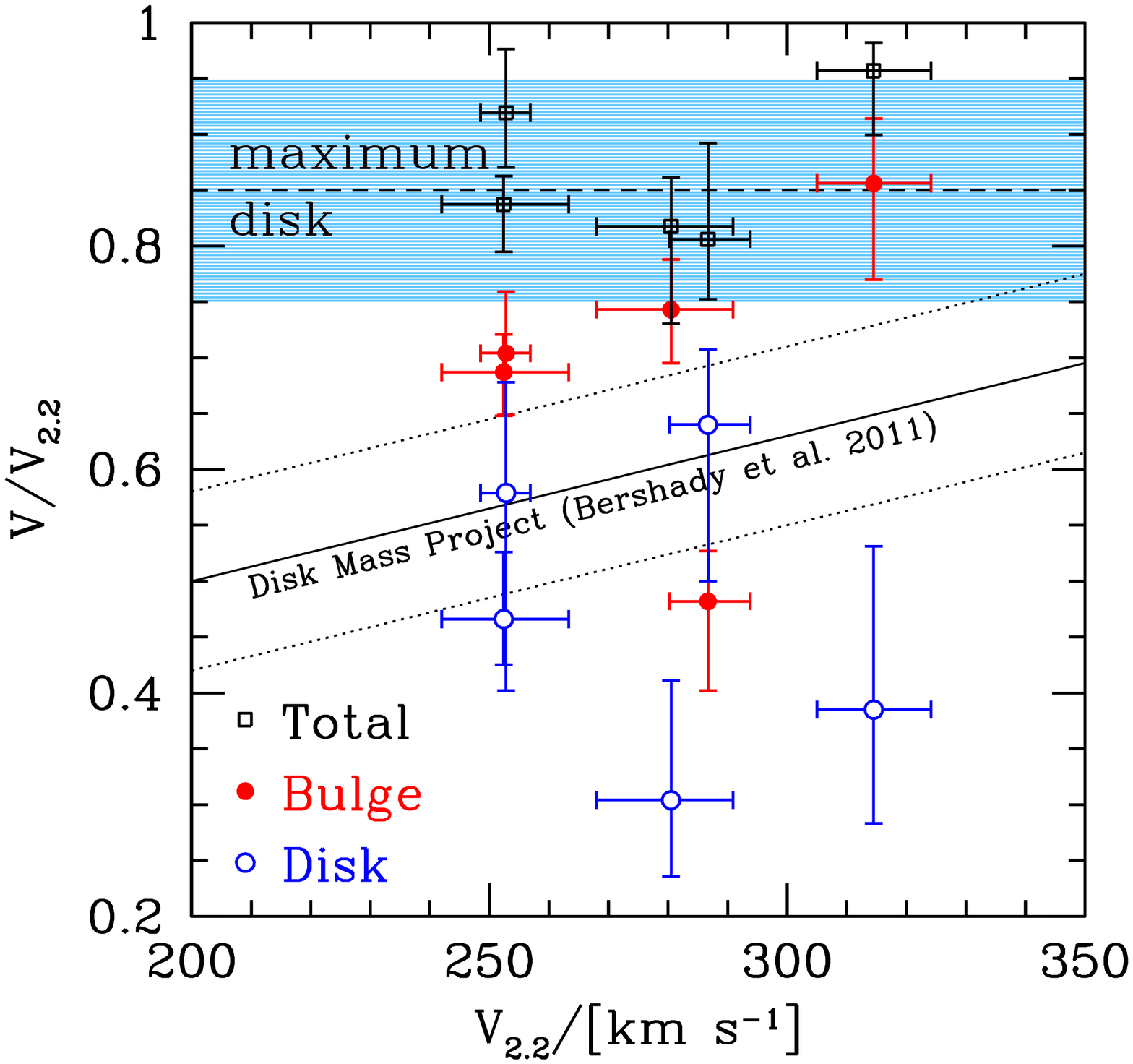}
\caption{Contribution of disk, bulge and bulge+disk to circular
  velocity at 2.2 disk scale lengths. Disks are sub-maximal in
  agreement with the Disk Mass Project \citep{Bershady2011}, where the
  solid and dashed lines correspond to their best fit relation and
  scatter, respectively.  However, the contribution of the disk plus
  bulge at 2.2 disk scale lengths is maximal, and thus our galaxies
  are baryon dominated within 2.2 disk scale lengths.}
\label{fig:fdisk}
\end{figure}

According to the standard definition \citep{Sackett1997}, a disk is
maximal if its contribution to the circular velocity at 2.2 disk scale
lengths is $0.75 < V_{\rm disk}/V_{2.2} < 0.95$. This is shown as a
blue shaded region in Fig.~\ref{fig:fdisk}. The blue circles in
Fig.~\ref{fig:fdisk} show that our five galaxies have $V_{\rm
  disk}/V_{2.2} < 0.7$, and are thus sub-maximal. This is in broad
agreement with results from the Disk Mass Project (solid and dashed
lines in Fig.~\ref{fig:fdisk}) which argued that all galaxy disks are
sub-maximal \citep{Bershady2011}, in addition to earlier studies
\citep[e.g.,][]{Bottema1993,CourteauRix1999}.

The galaxies in the Disk Mass Project are disk-dominated, so
sub-maximal disks imply their galaxies are not baryon dominated inside
2.2 disk scale lengths (i.e., there is significant dark
matter). However, in our galaxies there is a substantial bulge
contribution (usually more than the disk) to $V_{2.2}$. The
contribution of the stars (bulge plus disk) to $V_{2.2}$ is $\simeq
0.75$ to $0.95$ (black points in Fig.~\ref{fig:fdisk}), and thus these
galaxies can be considered maximal in their total baryonic content at
2.2 disk scale lengths. (See also the results of the analysis of a
disk galaxy lens including stellar kinematic constraints in SWELLS-IV,
\citealt{Barnabe2012}.)

\begin{figure*}
\centering
\includegraphics[width=0.48\linewidth]{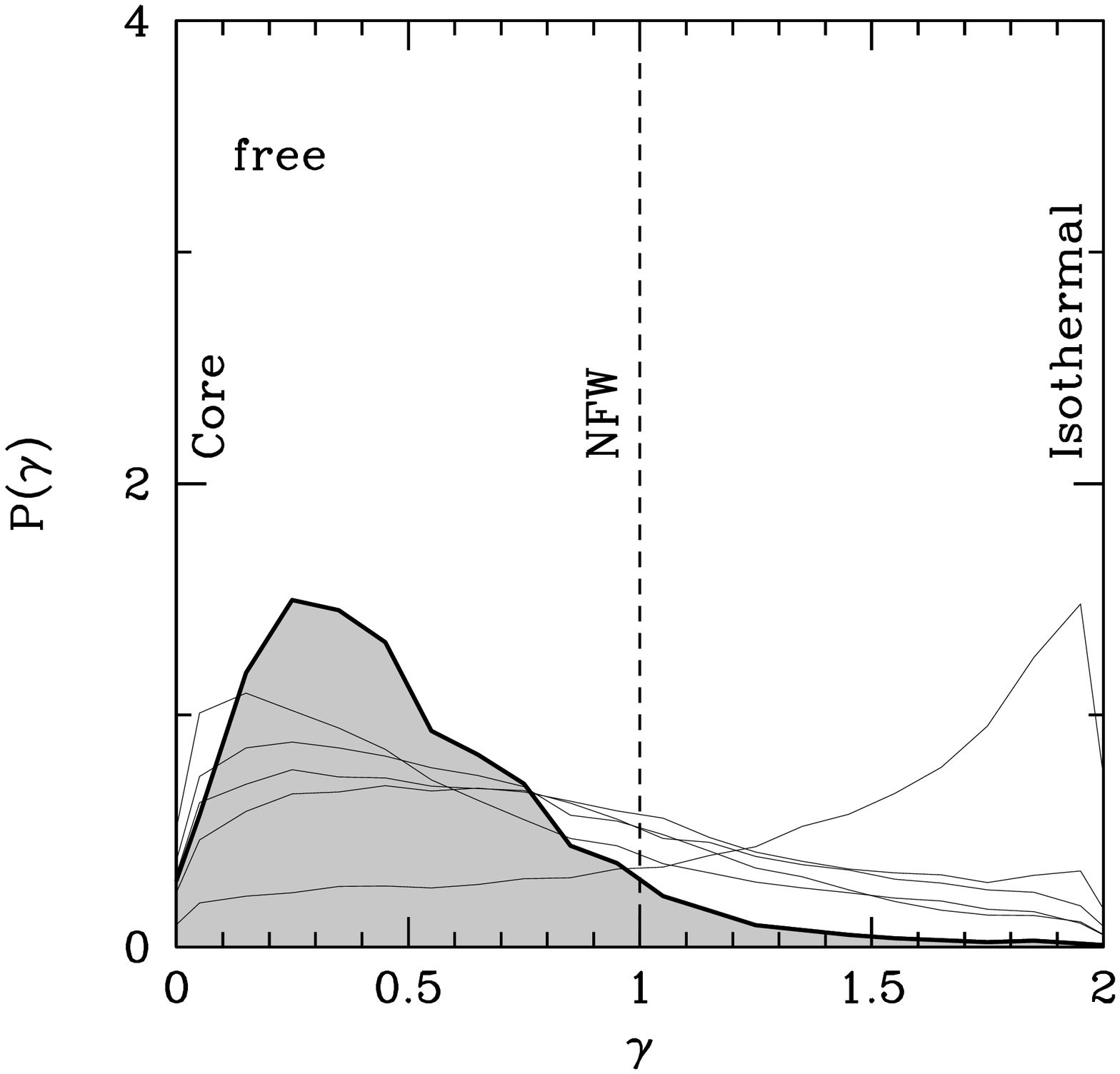}
\includegraphics[width=0.48\linewidth]{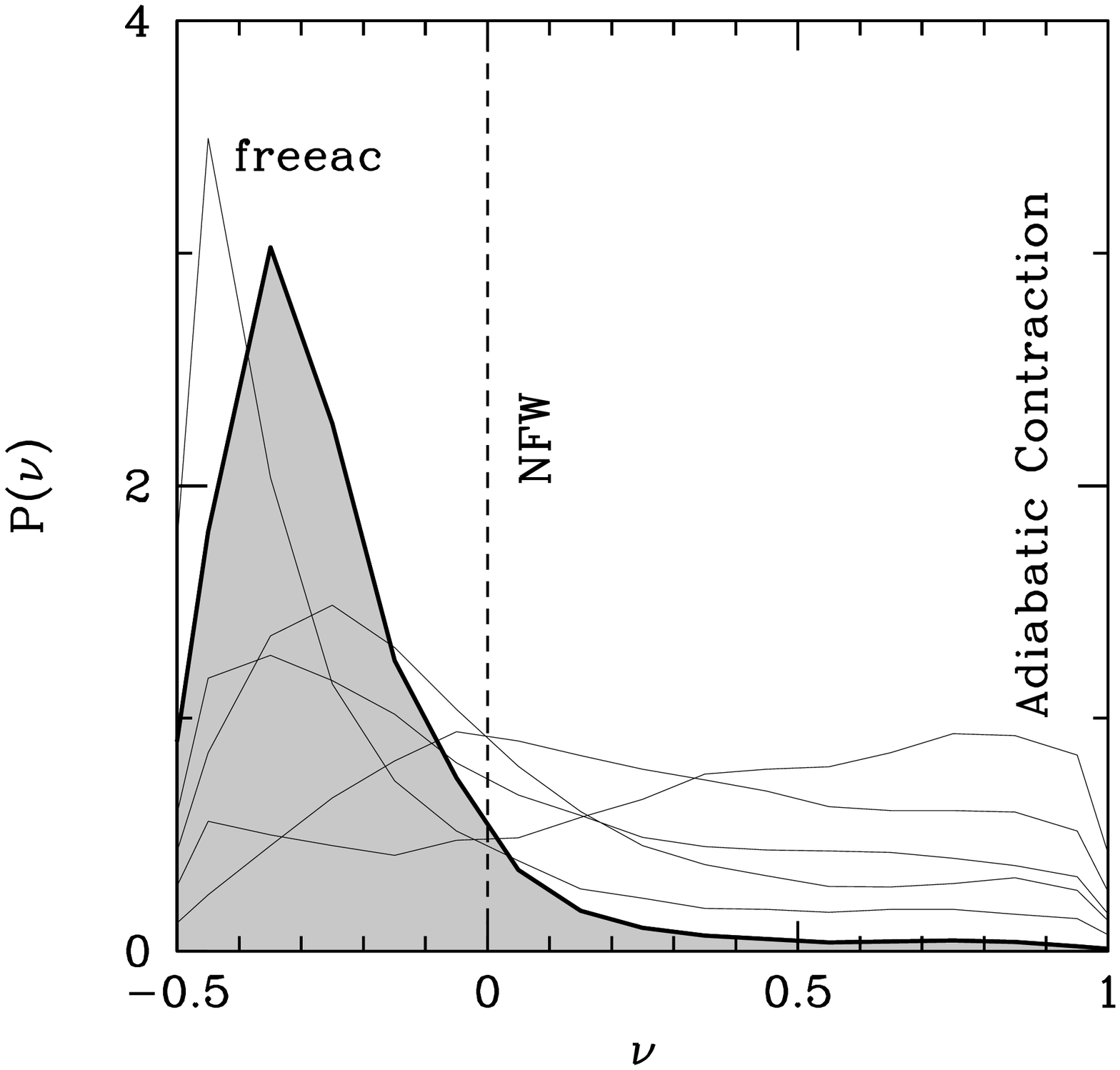}
\caption{Posterior distributions for the central slope of the dark
  matter density profile, $\gamma$ (left panel), and halo response
  parameter, $\nu$ (right panel). The bold histograms show the joint
  constraint on the mean. Haloes with $\gamma \gta 1$ or halo
  contraction $\nu > 0$ are, on average, disfavoured. }
\label{fig:gamma}
\end{figure*}

\subsection{Are dark matter haloes cuspy?}
The use of general dark matter halos allows us to investigate the
inner slope of the dark matter density profile, which is predicted by
numerical simulations to be approximately unity, in the absence of
baryons. Fig.~\ref{fig:gamma} shows the PDFs of $\gamma$ as inferred
from our data. For individual galaxies $\gamma$ is only weakly
constrained, reflecting the fact that the bulge stellar mass dominates
in the inner regions. Taken as an ensemble, and assuming a universal
value for $\gamma$, values larger than unity (i.e., NFW) are
marginally disfavoured [p($\gamma>1$)=0.07], consistent with what is
found for early-type galaxies of comparable mass
\citep{Treu-Koopmans2004, Dutton-Maccio-Mendel-Simard2012}, albeit
there are counter-examples \citep{Sonnenfeld2012,Grillo2012}, which
may indicate that there is a broad scatter stemming perhaps from
different formation histories.

At face value this result would imply a marginal conflict with the
universal profiles predicted by cold dark matter only numerical
simulations. However, since the central regions of these galaxies are
baryon dominated, an accurate comparison depends crucially on how
baryonic effects alter the underlying dark matter halos. Standard
arguments suggest that baryonic cooling to form stars should steepen
the overall mass density profile, thus causing the dark matter halo to
steepen as well \citep{Blumenthal1986, Gnedin2004}. This would only
exacerbate the tension between our observations and theoretical
prediction. However, in practice other processes may occur that act to
expand the halo, such as mass outflows due to stellar and/or active
galactic nuclei \citep[e.g.,][]{Read-Gilmore2005, Duffy2010,
  Pontzen-Governato2012} or dynamical friction between infalling
galaxies and the dark matter halo \citep[e.g.,][]{El-Zant2001,
  Johansson2009}.
Investigating this complex physics in detail goes beyond the scope of
this paper, but it is interesting to notice that our data suggest that
the net effect of baryonic physics appears to lead to real halos that
are flatter or at most as flat as NFW, and not as cuspy as the
standard contraction recipes would imply.

\begin{table}
\begin{center}
  \caption{Summary of priors on dark matter halo for our various mass
    models. $U(a,b)$ is a uniform prior with limits $a$, and
    $b$. $\delta(a)$ is a delta function prior at $a$.}
\label{tab:priors}
\scriptsize\begin{tabular}{lcccc}
  \hline
  \hline Model & $\gamma$ & $\nu$ & $r_{\rm s}$ & $V_{200}$\\
  \hline
  free & $U(0.0,2.0)$ & $\delta(0.0)$ & $r_{\rm s}(V_{\rm 200})$ & $U(V_{200}^{\rm min},V_{200}^{\rm max})$\\
  atlas& $U(0.0,1.6)$ & $\delta(0.0)$ & $\delta(20 {\rm kpc})$ & $U(V_{200}^{\rm min},V_{200}^{\rm max})$\\
  mfl  &  -           &  -            &  -                     &  - \\
  core & $\delta(0.0)$& $\delta(0.0)$ & $r_{\rm s}(V_{\rm 200})$ & $U(V_{200}^{\rm min},V_{200}^{\rm max})$\\
  nfw  & $\delta(1.0)$& $\delta(0.0)$ & $r_{\rm s}(V_{\rm 200})$ & $U(V_{200}^{\rm min},V_{200}^{\rm max})$\\
weakac & $\delta(1.0)$& see text & $r_{\rm s}(V_{\rm 200})$ & $U(V_{200}^{\rm min},V_{200}^{\rm max})$\\
  ac   & $\delta(1.0)$& $\delta(1.0)$ & $r_{\rm s}(V_{\rm 200})$ & $U(V_{200}^{\rm min},V_{200}^{\rm max})$\\
 freeac& $\delta(1.0)$& $U(-0.5,1.0)$ & $r_{\rm s}(V_{\rm 200})$ & $U(V_{200}^{\rm min},V_{200}^{\rm max})$\\
  \hline \hline
\end{tabular}
\medskip\\
\end{center}
\end{table}

\section{Testing systematic effects}
\label{sec:systematics}

In this section we discuss how our results depend on the functional
form of the dark matter halo (\S~\ref{ssec:amm}), and we test
potential systematic uncertainties arising from contamination of the
lensing signal by mass along the line of sight (\S~\ref{ssec:los}).
In \S~\ref{ssec:gas} we discuss briefly the effects of neglecting cold
gas in our lensing and dynamical analysis.  We find that our
conclusions are robust with respect to the choice of mass models and
line of sight effects, and the inclusion of cold gas.

\begin{figure*}
\centering
\includegraphics[width=0.24\linewidth]{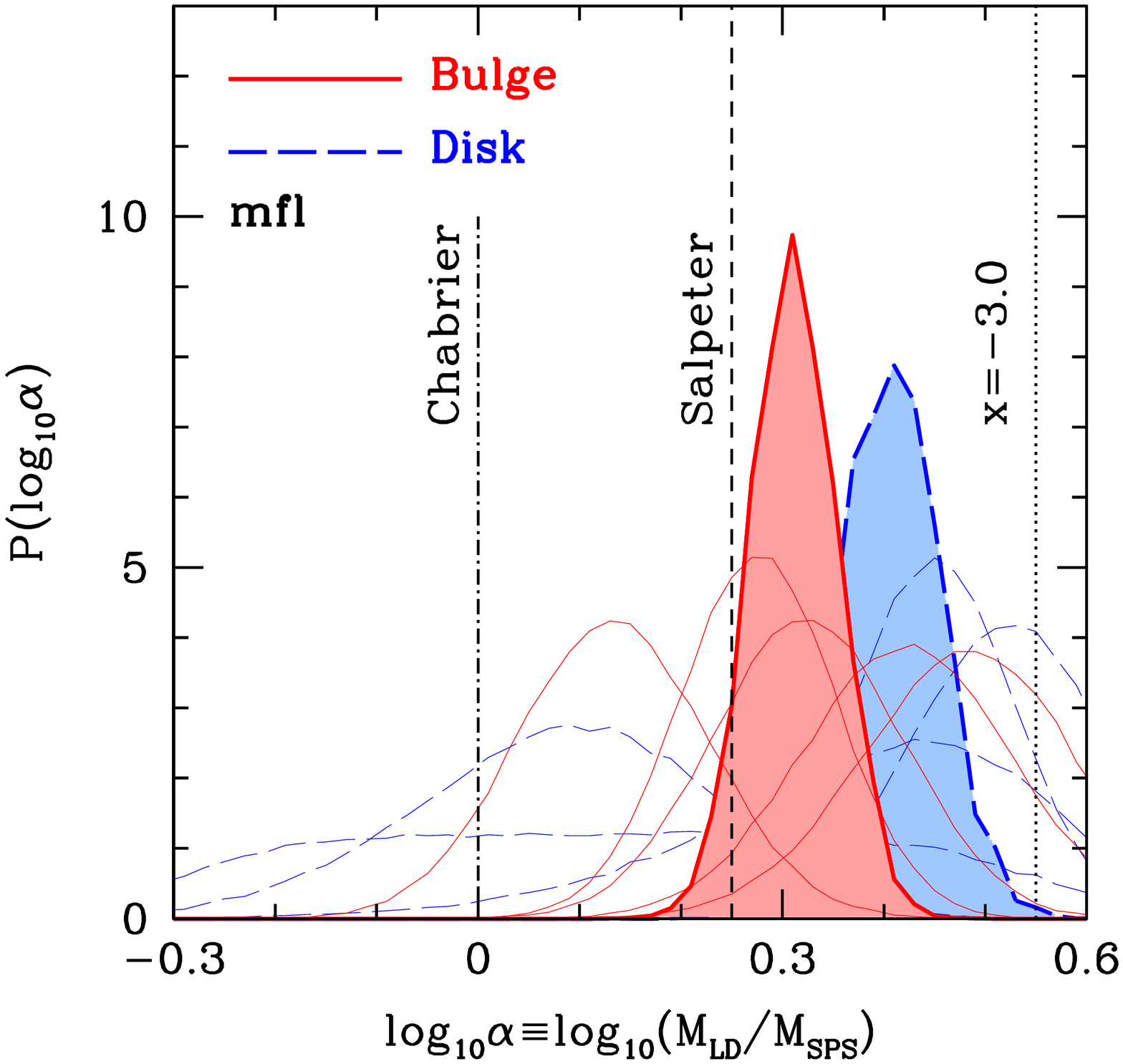}
\includegraphics[width=0.24\linewidth]{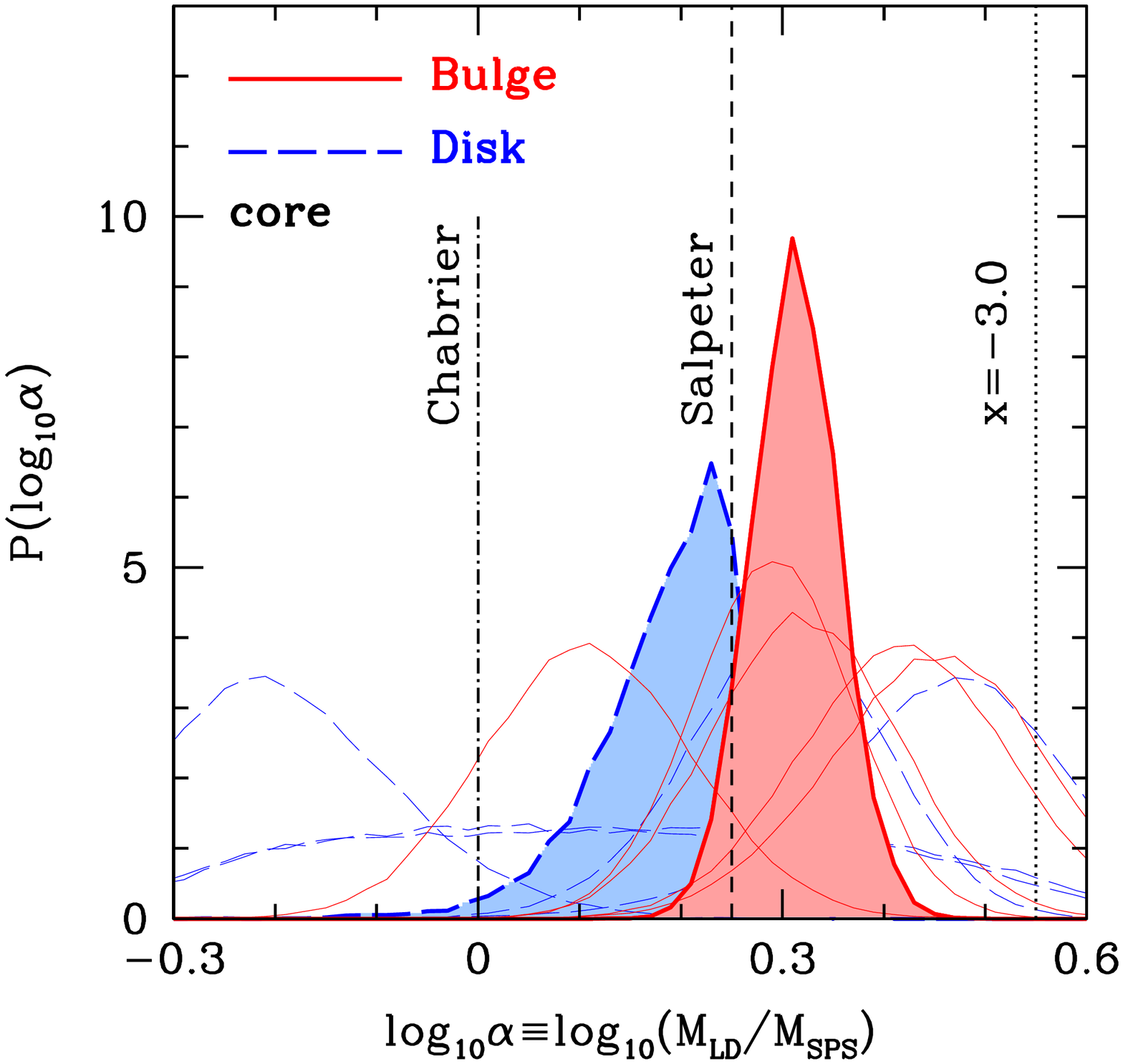}
\includegraphics[width=0.24\linewidth]{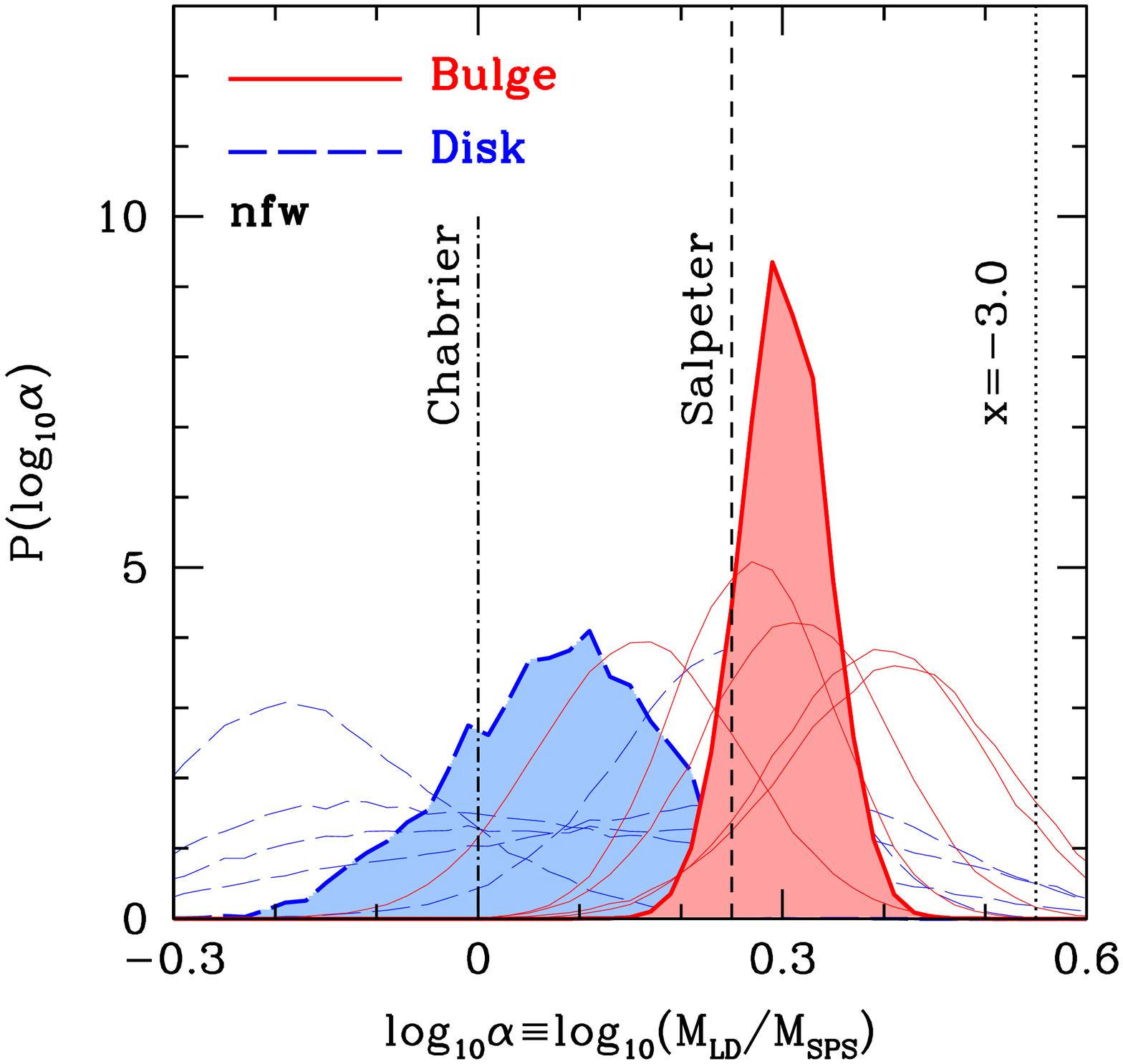}
\includegraphics[width=0.24\linewidth]{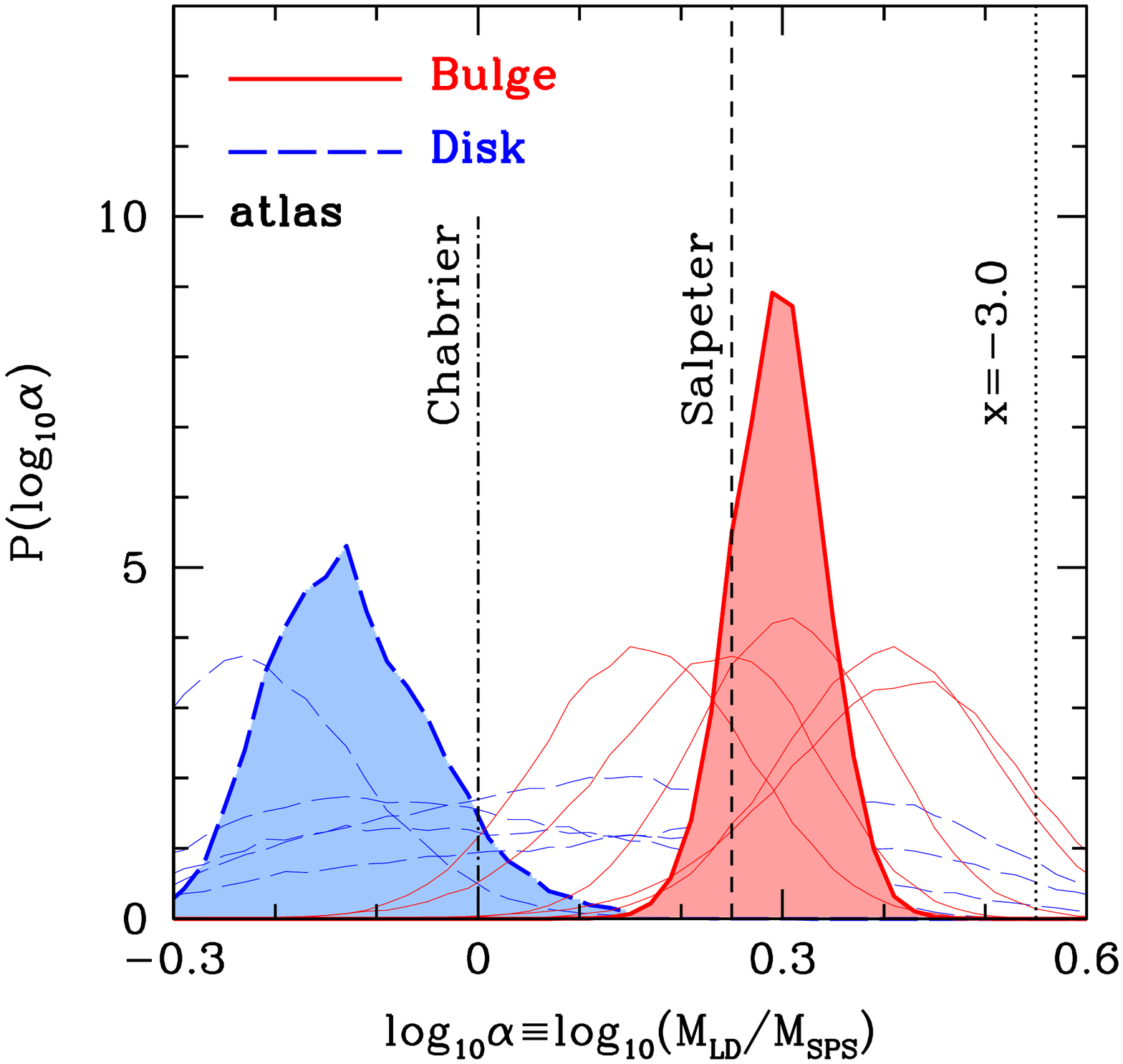}
\includegraphics[width=0.24\linewidth]{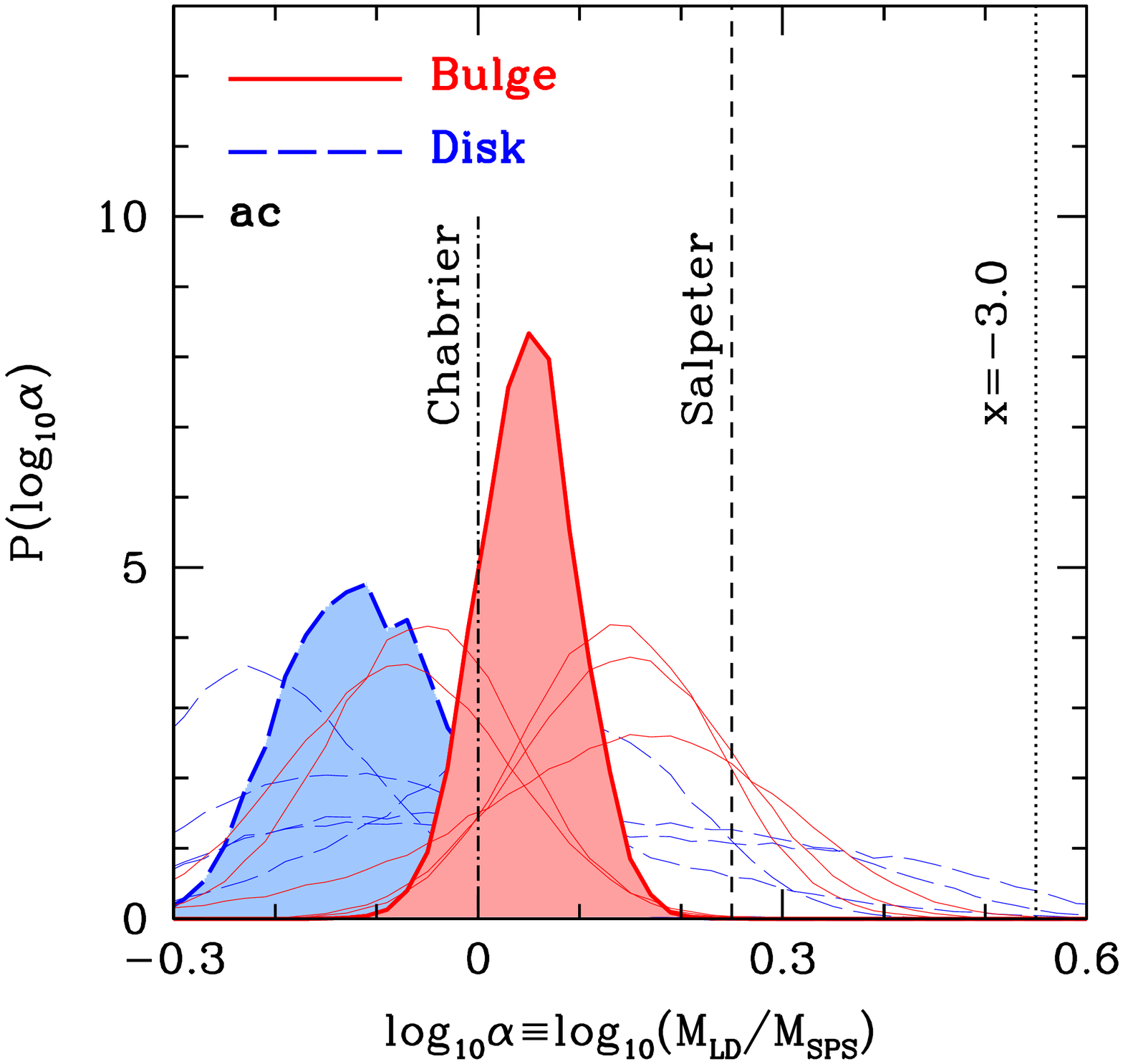}
\includegraphics[width=0.24\linewidth]{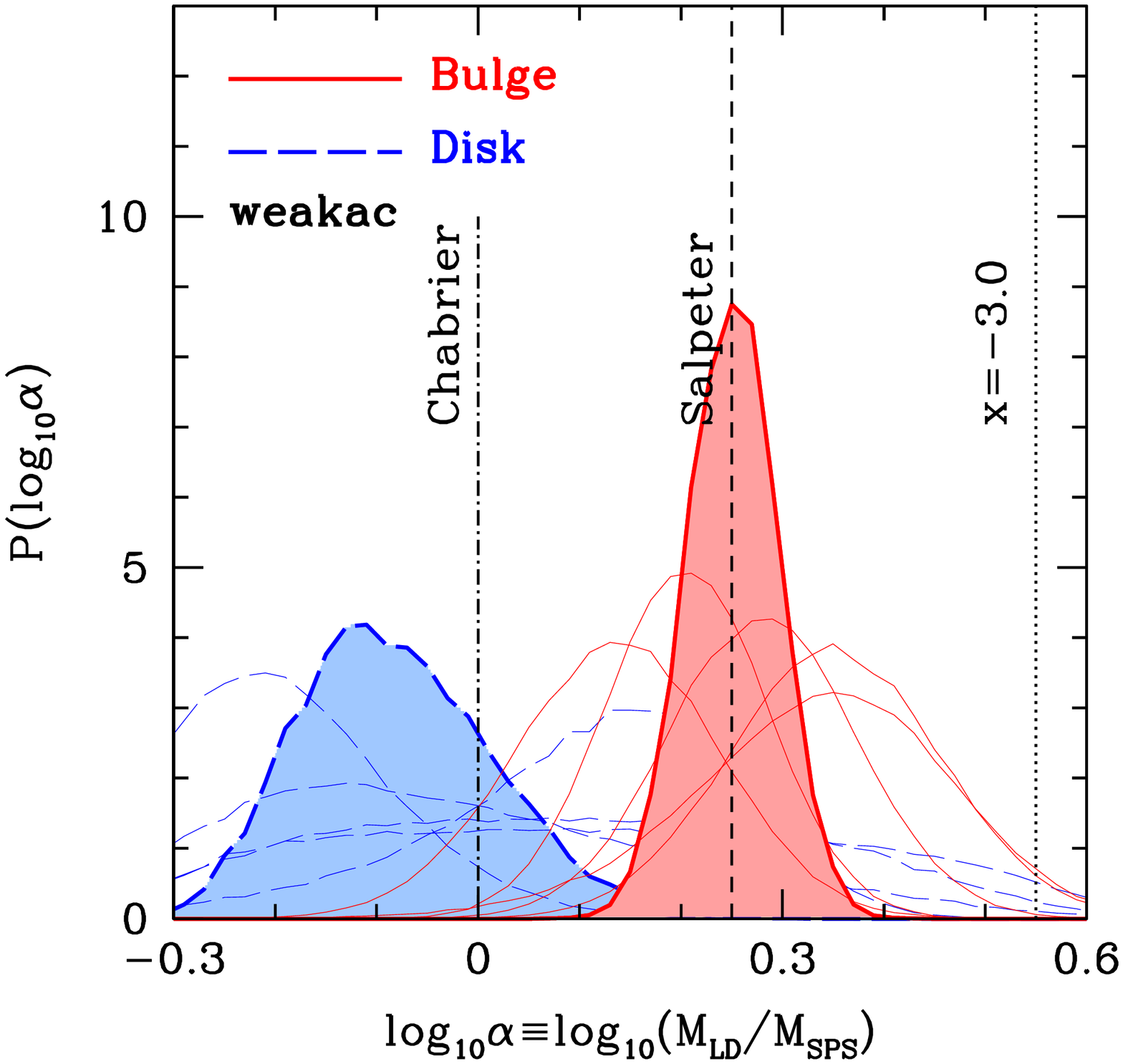}
\includegraphics[width=0.24\linewidth]{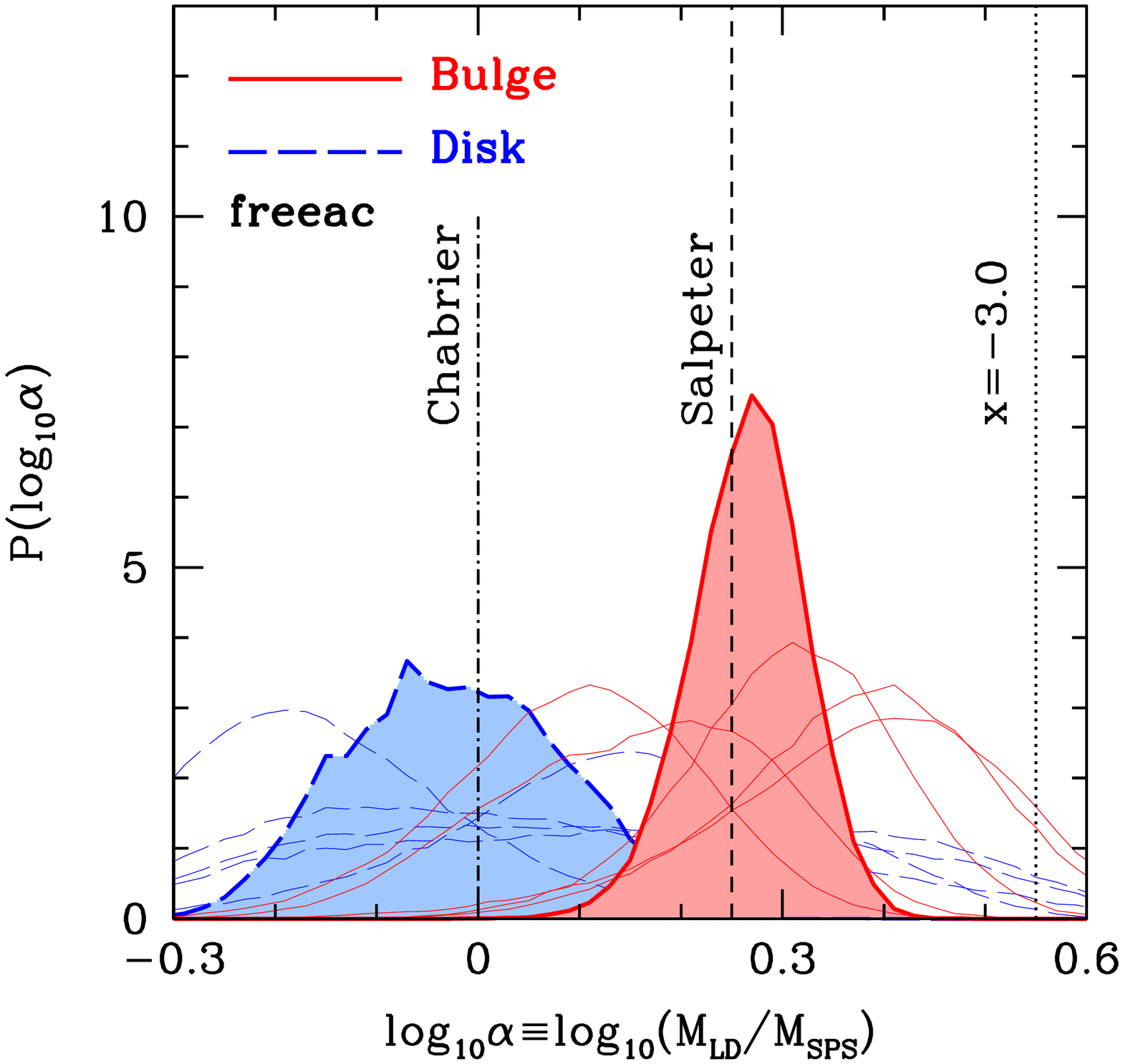}
\includegraphics[width=0.24\linewidth]{figs-bulges/alpha_hist_free9.ps}
\\
\includegraphics[width=0.24\linewidth]{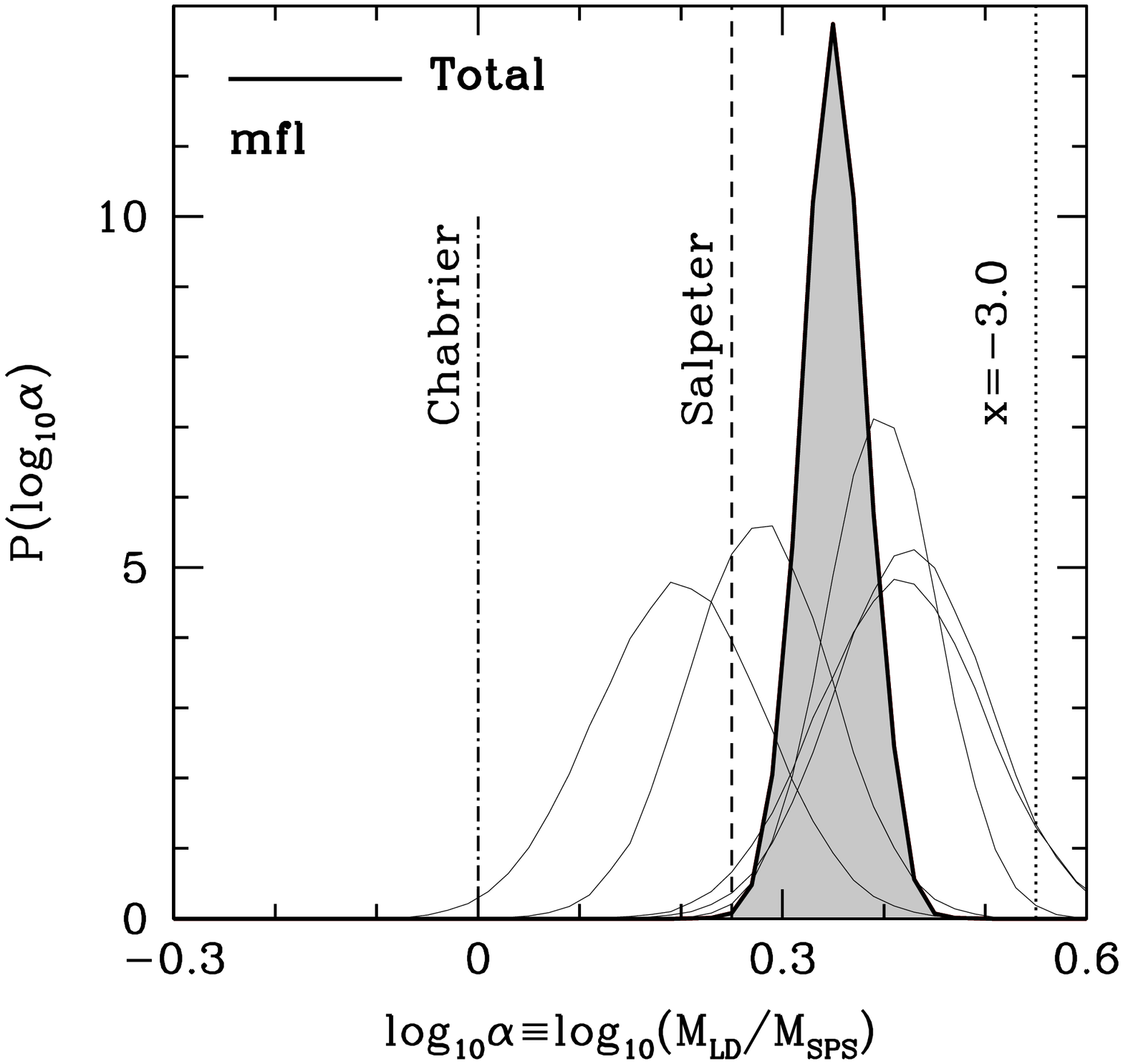}
\includegraphics[width=0.24\linewidth]{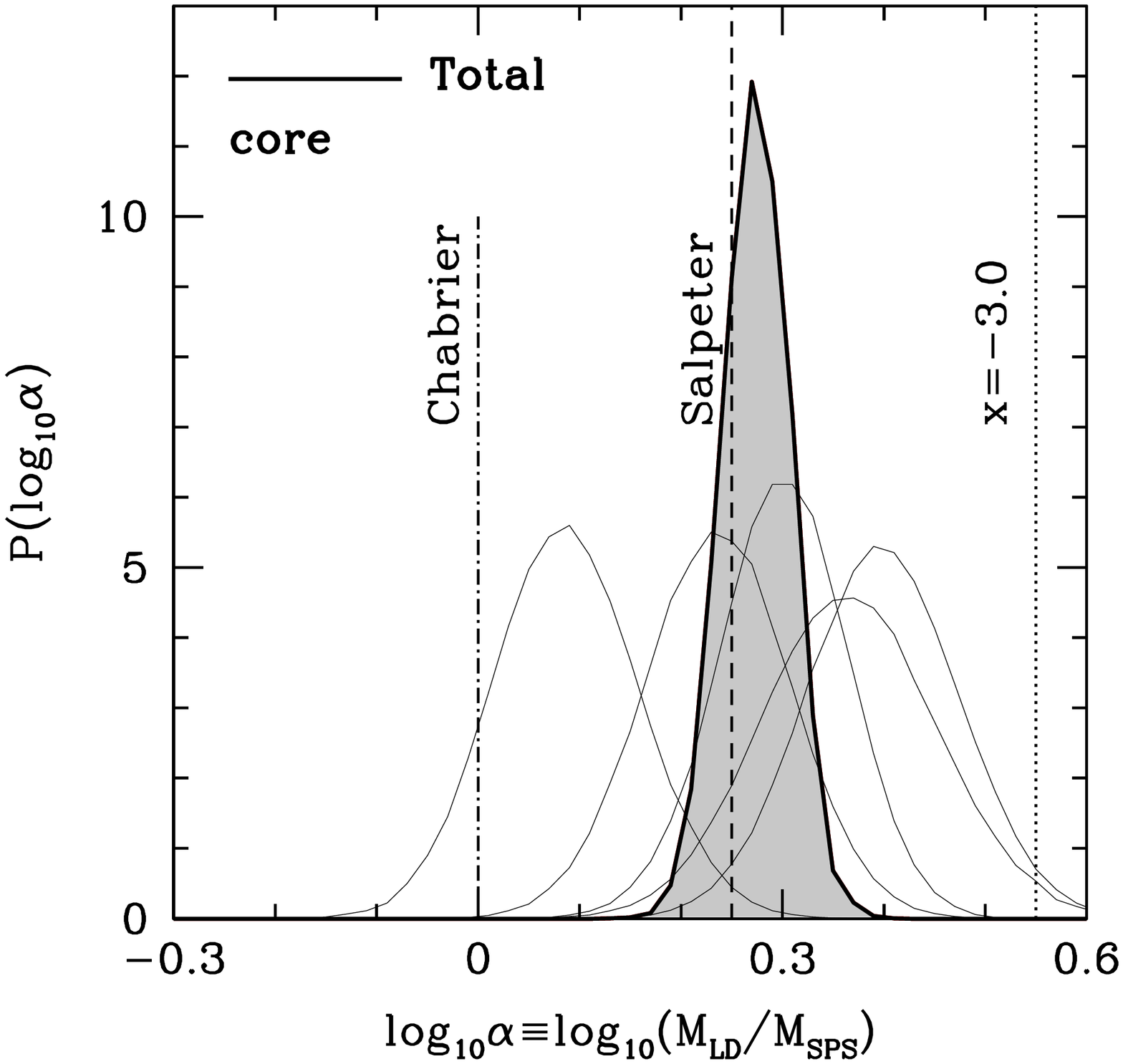}
\includegraphics[width=0.24\linewidth]{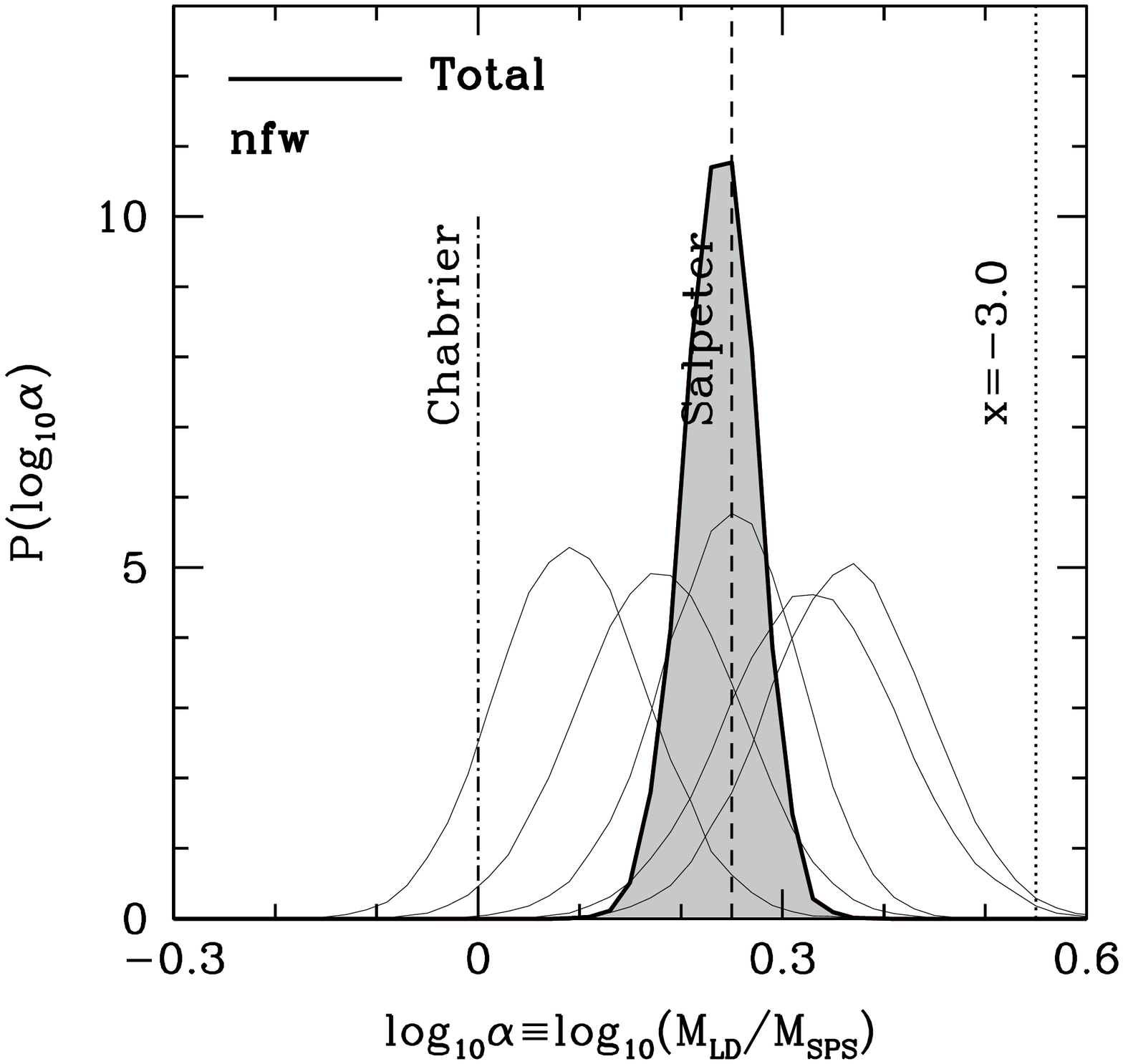}
\includegraphics[width=0.24\linewidth]{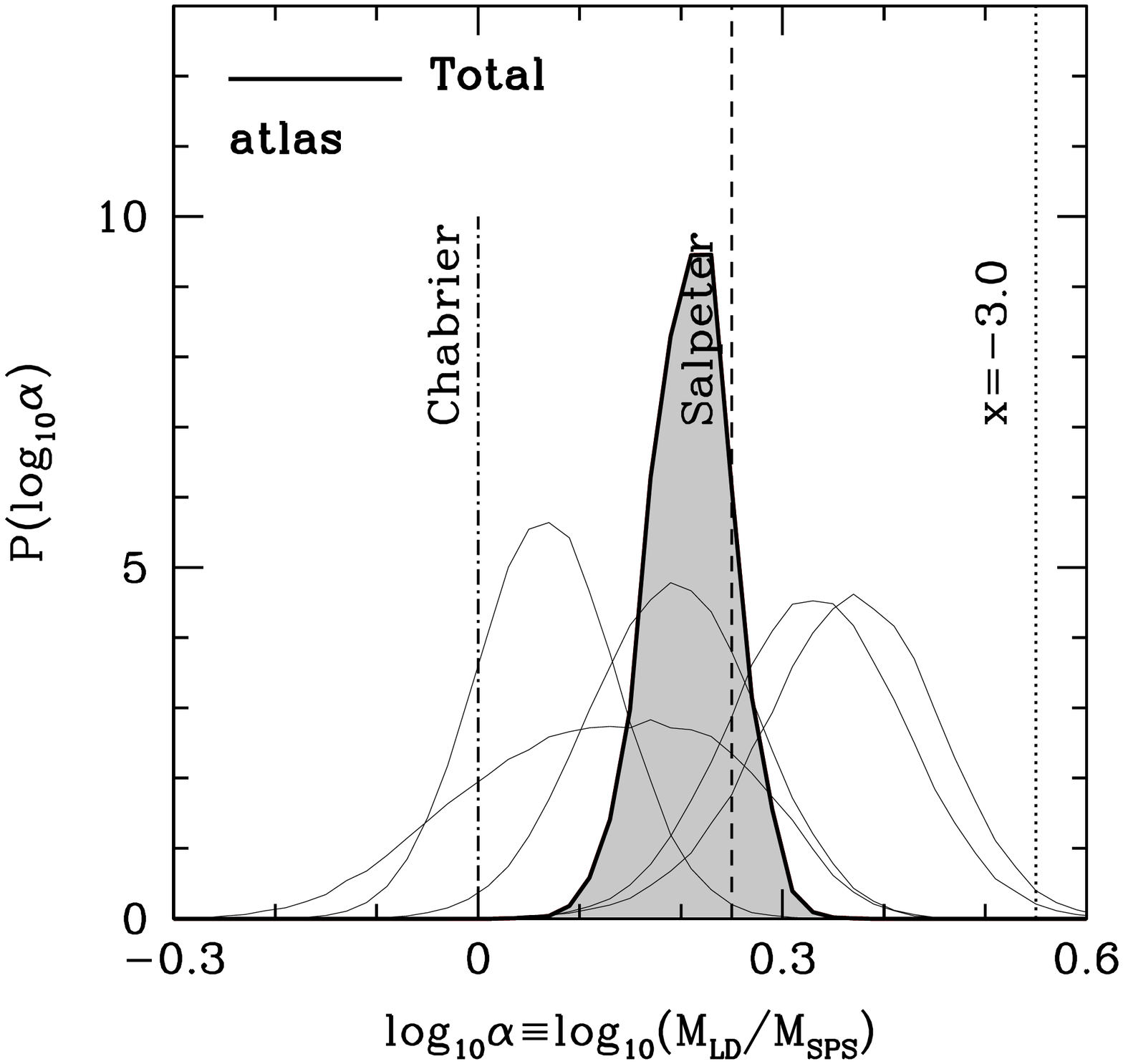}
\includegraphics[width=0.24\linewidth]{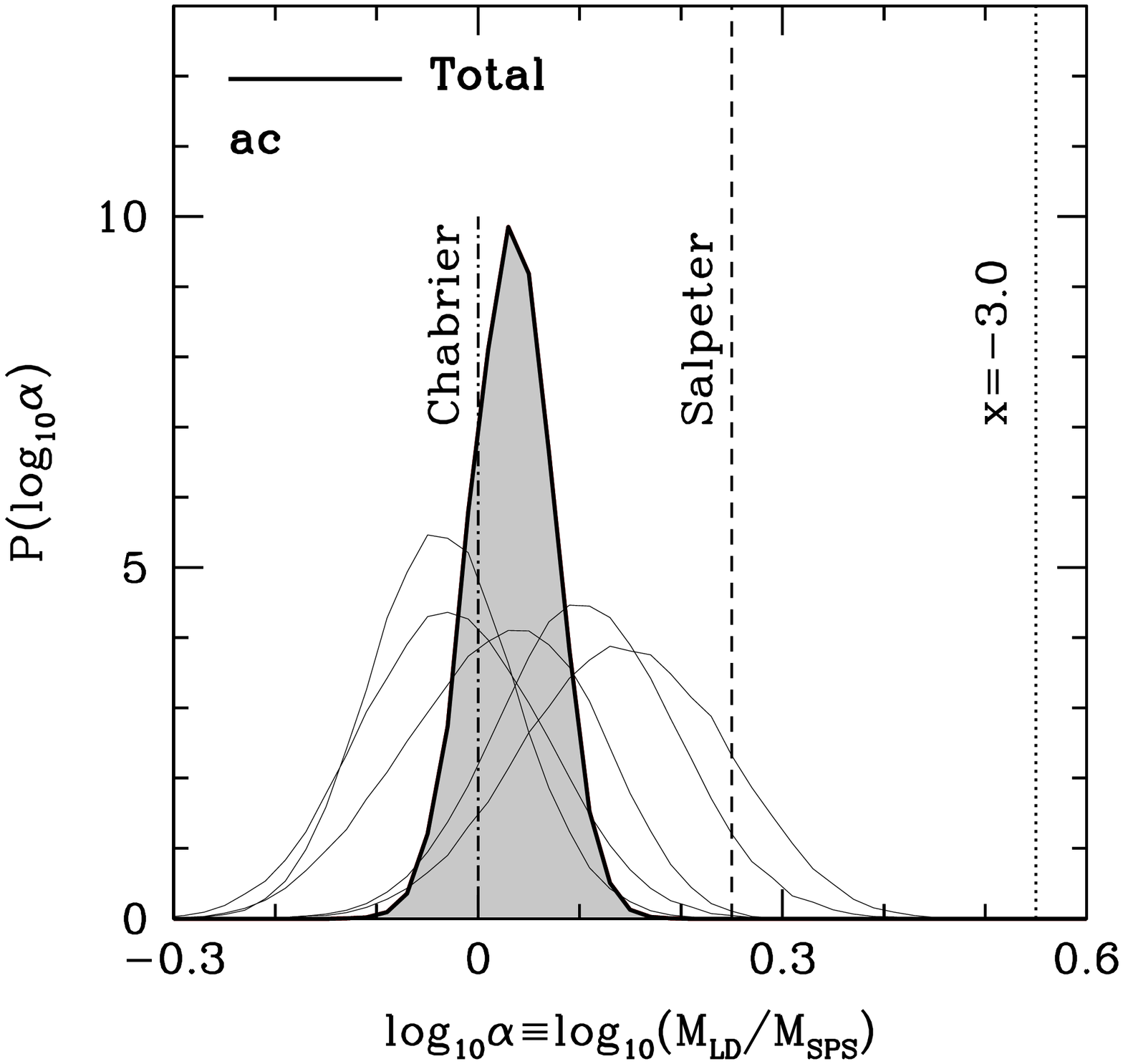}
\includegraphics[width=0.24\linewidth]{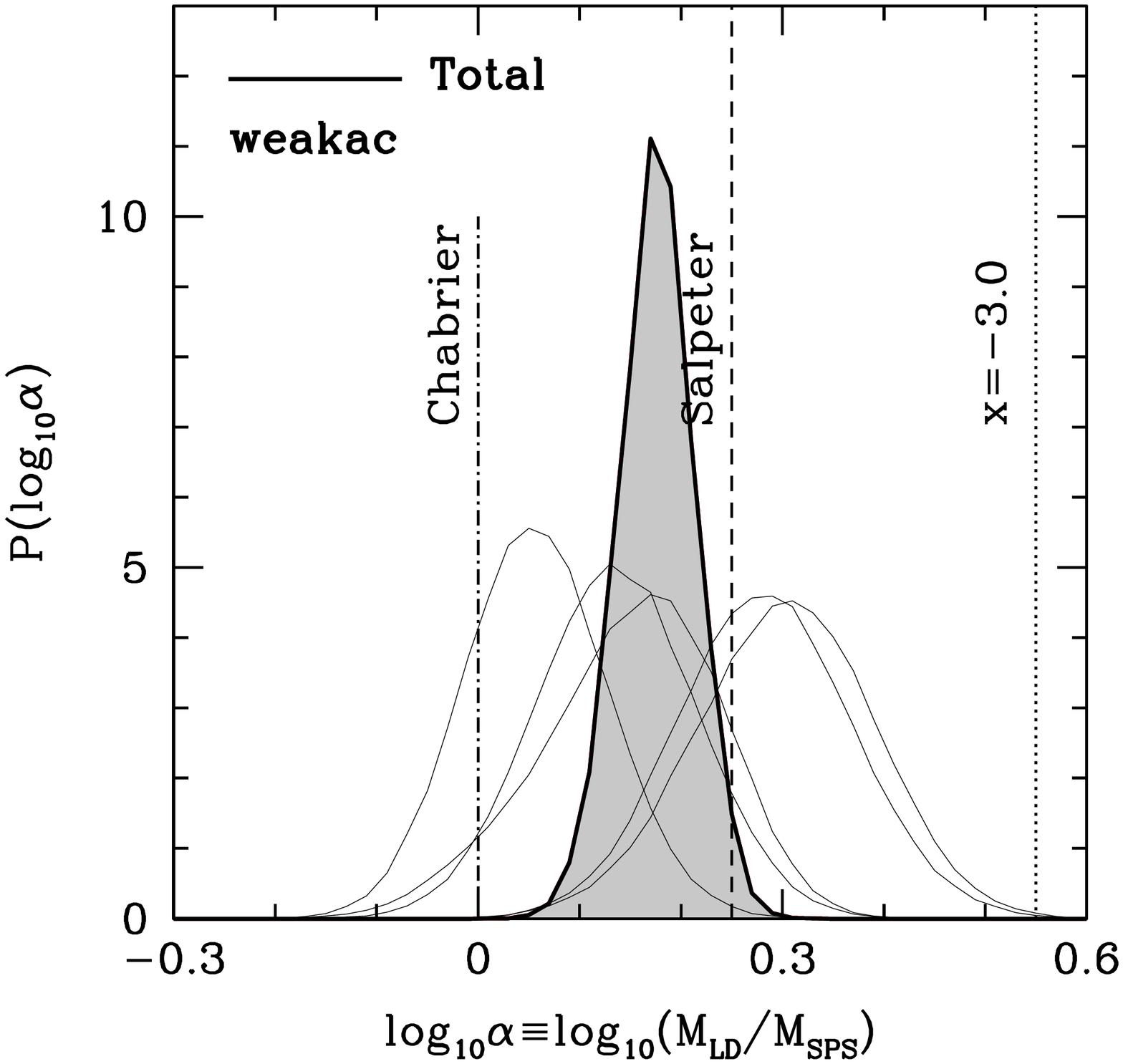}
\includegraphics[width=0.24\linewidth]{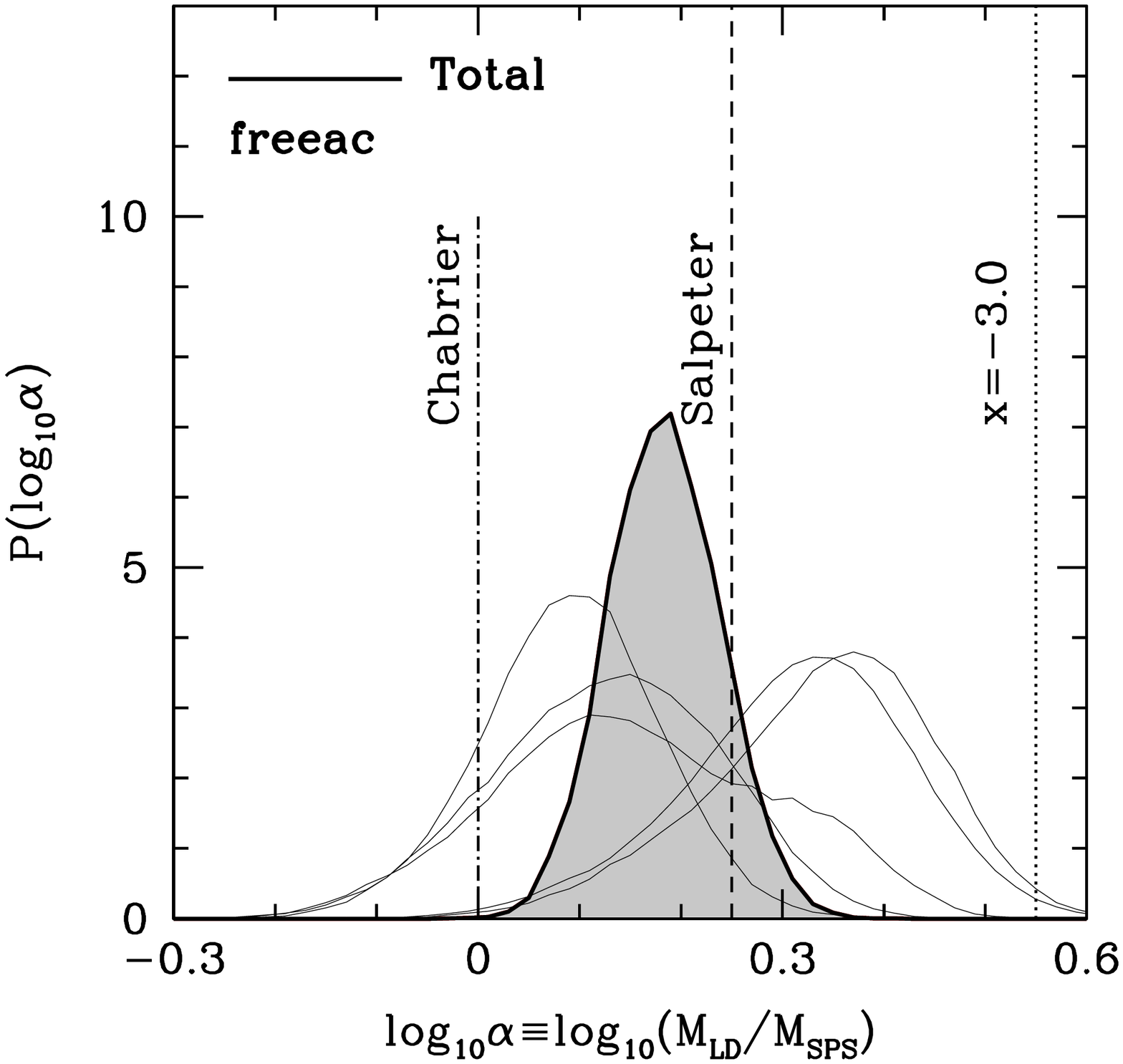}
\includegraphics[width=0.24\linewidth]{figs-bulges/alpha_hist_total_free9.ps}
\caption{Posterior distributions for the IMF mismatch parameter
  relative to a Chabrier IMF, $\alpha$, for eight different
  assumptions about the functional form of the dark matter halo: mfl
  --- mass-follows-light (i.e., no dark matter halo); core --- cored
  dark matter halo; nfw --- \citet{NFW1997} dark matter halo,
  constrained to follow the concentration-mass relation from
  \citet{Maccio2008} allowing for scatter; ac --- adiabatically
  contracted nfw model; weakac --- nfw model with weak halo
  contraction following \citet{Abadi2010}; freeac --- nfw model with
  free halo response ranging from adiabatic contraction to expansion;
  atlas --- halo profile adopted by ATLAS3D \citep{Cappellari2012}
  which is a generalised NFW with fixed scale radius; free ---
  generalised NFW halo with free inner slope (repeated from
  Fig.~\ref{fig:alpha}).  {\it Upper:} $\alpha$ for bulges (red, solid
  lines) and disks (blue, long dashed lines); {\it Lower:} $\alpha$
  for total (bulge+disk) stellar mass. The bold histograms show the
  joint constraints on the mean $\alpha$. For all models except ``ac''
  the bulges are consistent with a Salpeter IMF. For ``ac'' the bulges
  are consistent with a Chabrier IMF, but allowing the halo response
  to vary ``freeac'' generally finds better fits with weaker halo
  contraction or expansion.}
\label{fig:alpha_all}
\end{figure*}

\subsection{Alternative dark matter halo models}
\label{ssec:amm}

Our inferred dark matter profile is less cuspy than the NFW profile
and one may be concerned that we have fitted away part of the central
dark matter into our bulge component, thus leading to the preference
for Salpeter-like IMFs over Chabrier. We therefore also consider a
wide range of less flexible dark matter haloes, ranging from models
with no dark matter or ``mass-follows-light'' (``mfl''), to cored dark
matter haloes (``core''), to contracted NFW haloes (``ac'',
``weakac'', and ``freeac''). The most generic model adopted by the
ATLAS3D team in their recent study \citep{Cappellari2012} is also
included to facilitate comparisons between our works (``atlas'').

The contracted NFW halos are meant to represent the response of the
dark matter halo to galaxy formation. The simple model ``ac'' refers
to the standard \citet{Blumenthal1986} formalism, the ``weakac'' model
is the reduced contraction model of \citet{Abadi2010}, while the
``freeac'' model is a generalisation.  In the standard formalism the
adiabatic invariant is $rM(r)$, where $r$ is galactic radius and
$M(r)$ is the spherically enclosed mass within $r$. Thus
\begin{equation}
\label{eq:ac}
r_{\rm f}/r_{\rm i}=M_{\rm i}(r_{\rm i})/M_{\rm f}(r_{\rm f}),
\end{equation}
where $r_{\rm i}$ and $r_{\rm f}$ are the initial and final radii,
respectively.  

In order to explore the possibility of weaker halo contraction
\citep[e.g.,][]{Abadi2010} and even expansion, we also consider the
generalised contraction formula from \citet{Dutton2007}. If the
standard contraction ratio from Eq.~\ref{eq:ac} is $\Gamma=(r_{\rm
  f}/r_{\rm i})$, then the modified contraction ratio is given by
$\Gamma^{\nu}$. Standard adiabatic contraction \citep{Blumenthal1986}
corresponds to $\nu=1$, the \citet{Gnedin2004} model can be
approximated with $\nu\sim 0.8$, the \citet{Abadi2010} model can be
approximated with $\nu\sim 0.4$, no contraction corresponds to
$\nu=0$, while expansion corresponds to $\nu < 0$.

Table ~\ref{tab:priors} lists the dark halo priors for the eight models
we consider here. We note that the models ``core'' and ``nfw'' are
specific realisations of the model ``free'', and the models ``ac'' and
``nfw'', are specific realisation of the model ``freeac''. 

The posterior probability distribution functions for the IMF mismatch
parameters of the bulge, disk, and total stellar mass for all models
(our reference ``free'' model and alternatives) are given in
Fig.~\ref{fig:alpha_all}.  As expected, the disk masses are somewhat
sensitive to the choice of dark matter halo. However, the bulge masses
are remarkably insensitive to the dark matter halo model. All of the
models, with the exception of adiabatically contracted NFW haloes
(model ``ac''), favour Salpeter-type IMFs for the bulges. The ``ac''
model is the only one that favours a Chabrier-type IMFs for the
bulges.  However, the fact that the ``freeac'' model contains the
``ac'' model allows us to perform a clean model selection procedure
and quantify whether the models with Salpeter IMF or the ``ac'' model
provide a better overall description of the data. Qualitatively, as
can be seen from Fig.~\ref{fig:gamma}, the ``freeac'' model prefers
values of the contraction index that disfavour the standard adiabatic
contraction models. Quantitatively the comparison between the
``freeac'' model and the ``ac'' model is given by the evidence ratio
\citep{Sivia-Skilling2006}. The evidence of the ``freeac'' model is 35
times larger than that of the ``ac'' model'', which corresponds to
strong evidence that the first model is to be preferred according to
standard criteria \citep{Sivia-Skilling2006}. Similar considerations
can be made for the IMF mismatch parameter $\alpha$ of the disks,
although in general it is less well constrained than that of the
bulge. All models with dark matter prefer disks that are Chabrier or
lighter, except for the ``nfw'' and ``core'' models. Those are subsets
of the ``free'' model and are disfavoured by the evidence ratios.

We thus conclude that {\it it is possible} to reconcile the data with
a Chabrier IMF for bulges if one asserts that standard adiabatic
contraction is the way dark matter halos respond to galaxy formation.
However, cosmological simulations of galaxy formation often find much
weaker contraction than predicted by the adiabatic contraction
formalism \citep{Johansson2009, Abadi2010, Tissera2010}, or even
expansion \citep{Governato2010, Maccio2012} and thus we do not
actually expect adiabatic contraction to occur in nature.
Furthermore, under the more general assumption of ignorance about the
effects of galaxy formation on the underlying dark matter halo, the
data clearly prefer uncontracted or mildly expanded halos with a
Salpeter IMF. This is true for a broad range of models, including our
``free'' models, our ``freeac'' models, and the ``atlas'' models.

\begin{figure}
\centering
\includegraphics[width=0.45\linewidth]{figs-bulges/alpha_hist_free9.ps}
\includegraphics[width=0.45\linewidth]{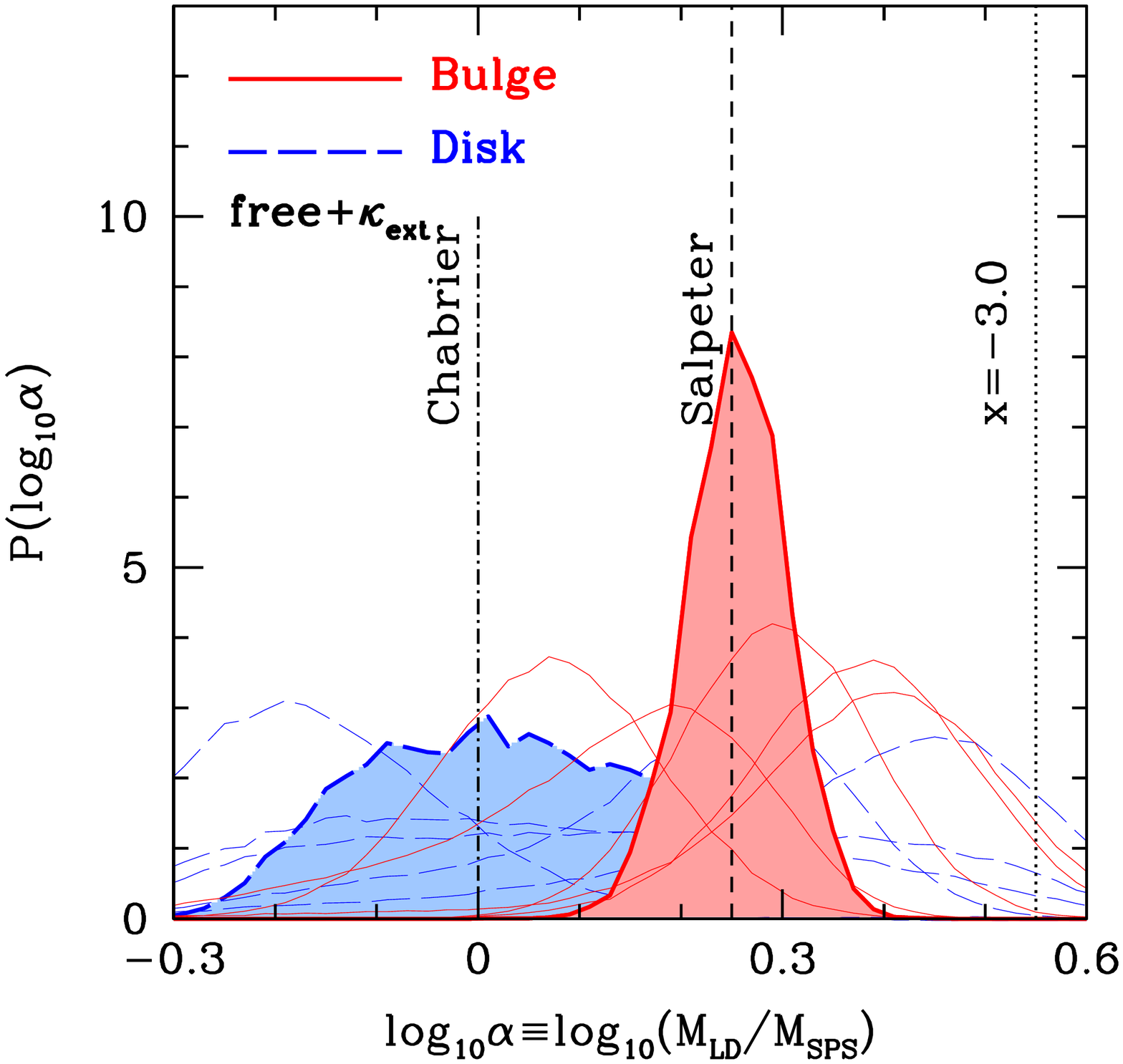}
\\
\includegraphics[width=0.45\linewidth]{figs-bulges/alpha_hist_total_free9.ps}
\includegraphics[width=0.45\linewidth]{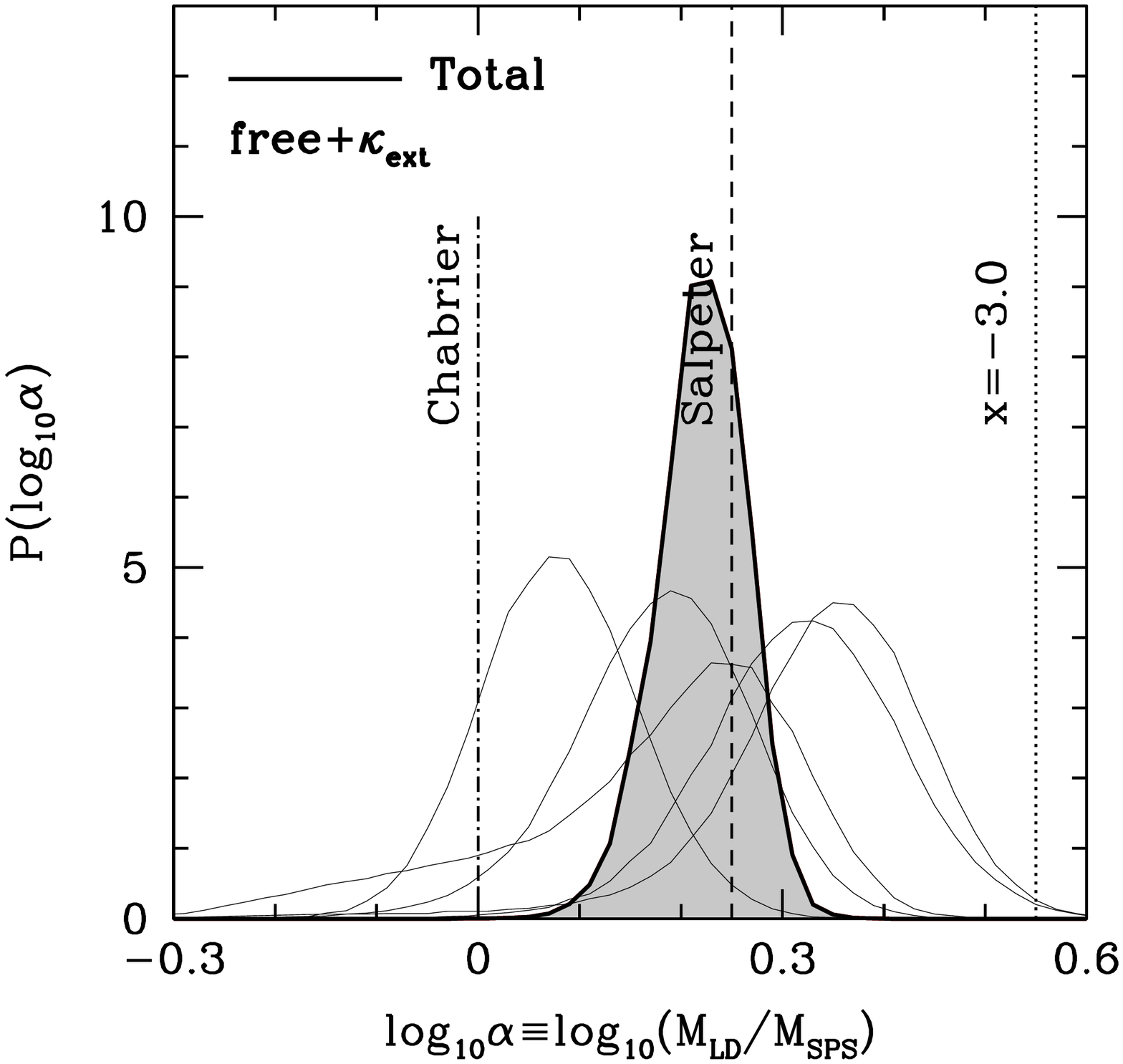}
\caption{Effect of external convergence on the posterior distributions
  for the IMF mismatch parameter relative to a Chabrier IMF, $\alpha$,
  for the ``free'' model. The model ``free$+\kappa_{\rm ext}$'' (right
  panels) has the same dark matter halo as ``free'' (left panels,
  repeated from Fig.~\ref{fig:alpha}), except we have included 5\%
  external convergence. The effect of external convergence is to lower
  $\alpha$ by a small amount.}
\label{fig:sigma_ext}
\end{figure}

\subsection{Line of sight effects}
\label{ssec:los}

Another potential concern is whether excess mass along the line of
sight could lead us to overestimate the mass associated with the main
deflector and thus overestimate $\alpha$. This is usually described in
the lensing literature as external convergence \citep[$\kappa_{\rm
  ext}$, e.g.,] []{Treu2010,Suyu2010}. The excess/deficit mass along
the line of sight acts to first approximation as a sheet of mass at
the redshift of the deflector, expressed in units of the critical
density $\kappa_{\rm crit}$, and cannot be measured with pure lensing
arguments due to the well known mass-sheet degeneracy
\citep{Falco1985}.  External convergence affects the lensing
observable used in our analysis -- the amount of mass within the
Einstein Radius -- in a very simple manner. The true mass will just be
the observed mass obtained by assuming the line of sight has the
average density of the universe, multiplied by (1-$\kappa_{\rm ext}$),
for small values of $\kappa_{\rm ext}$.

At the relatively low redshift of the SWELLS and SLACS samples,
external convergence is very small. Detailed analysis of the SLACS
sample shows that the external convergence is typically a few percent
\citep[e.g.,][]{Treu2009, Guimaraes-Sodre2011, Sonnenfeld2012}. Thus,
we expect that external convergence will change our results by a few
percent at most. For completeness, we repeated all our inference
assuming $\kappa_{\rm ext}=0.05$. As expected, $\alpha$ shifts down by
a negligible amount (compare model ``free'' to ``free$+\kappa_{\rm
  ext}$'' in Fig.~\ref{fig:sigma_ext}) demonstrating that our
inferences are robust with respect to line of sight effects.

\subsection{Cold Gas}
\label{ssec:gas}
In principle the stellar masses derived from our lensing plus dynamics
fits are upper limits, because we do not include cold gas in our mass
models. For the stellar masses of our galaxies we expect cold gas
fractions of $\sim 20\pm10\%$ \citep[e.g.,][]{Dutton2011-IMF}, roughly
equally split between atomic and molecular gas. Molecular gas
generally traces the stars, and is typically less than $\sim 10\%$ of
the stellar mass, assuming a Chabrier IMF \citep{Saintonge2011}, so
the effect on our derived bulge stellar masses is expected to be less
than 0.05 dex. The atomic gas typically has a larger scale length than
the stellar disk, so we expect the atomic gas to subtract mass from
the stellar disk and dark matter halo. In summary, we expect that
including observations of cold gas in our mass models will not
significantly reduce the derived stellar mass of the bulges.


\section{Summary}
\label{sec:concl}

We have presented mass models of 5 massive galaxies selected from the
SWELLS survey \citep{Treu2011} to have bulges and disks of comparable
stellar mass, as well as star forming disks.  We combined masses from
strong lensing with ionised gas kinematics at $\sim 1-3$ disk scale
lengths to constrain the parameters of three component mass models
consisting of a bulge, a disk and a generic dark matter halo. Our main
results can be summarised as follows:

\begin{itemize}

\item The stellar masses of the bulges are well constrained by the lensing and
  kinematic data, independent of the inner density slope of the dark
  matter halo.

\item The bulge masses inferred from the lensing and dynamical models,
  $M_{\rm LD}$, are inconsistent with those obtained from the colours,
  $M_{\rm SPS}$, assuming a Chabrier IMF, but in good agreement with those
  based on a Salpeter IMF.  The average normalisation of the IMF of
  the bulges relative to that based on a Chabrier IMF, is given by the
  IMF mismatch parameter $\log_{10}\alpha_{\rm bulge}\equiv
  \log_{10}(M_{\rm LD,bulge}/M_{\rm SPS,bulge}^{\rm
    Chab})=0.29\pm0.05$

\item The disk masses inferred from the lensing and dynamical models
  are only weakly constrained, due to degeneracies with the dark
  matter halo, but are consistent with a Chabrier-like IMF. The
  average IMF mismatch parameter is found to be
  $\log_{10}\alpha_{\rm disk}=-0.01\pm0.12$.

\item Disks are sub-maximal at 2.2 disk scale lengths (in agreement
  with the Disk Mass project, \citealt{Bershady2011}). However,
  baryons dominate the potential inside 2.2 disk scale lengths due to
  the strong bulge components. And thus sub-maximal disks do not imply
  galaxies are dark matter dominated inside 2.2 disk scale lengths.

\item The data marginally disfavour an inner slope of the dark matter
  halo $\gamma>1$ that would be expected for NFW halos contracted
  according to standard adiabatic contraction
  prescriptions. Equivalently, the data favour an uncontracted or
  marginally expanded NFW halo.

\end{itemize}

Our main new result is that IMF of bulges of spirals is ``heavier''
than Chabrier, and consistent with a Salpeter IMF.
Since our data do not strongly constrain the IMF of the disks, there
are two possible implications. Either the IMF varies {\it between}
spiral galaxies (e.g., massive spiral galaxies have heavier IMFs than
the Milky Way), or the IMF varies {\it within} spiral galaxies (e.g.,
bulges have heavier IMFs than disks). Since all previous constraints
on the masses of galactic disks seem to rule out Salpeter IMFs
\citep[e.g.,][]{Bell-deJong2001, Bershady2011, Dutton2011-IMF,
  Martinsson2011PhD, Barnabe2012} we favor the latter hypothesis.
Even though this result might seem surprising, it is well known that
the stellar populations of bulges and disks differ in age and chemical
composition. Thus, if the IMF reflects the physical conditions at the
time of formation of the stellar populations, it is entirely possible
that it could be different for the bulge and disk. Clearly, even
though this is possible, the underlying physical reasons are at
present unclear. Our hope is that this new piece of the puzzle will be
a valuable clue for deciphering the mystery of the IMF and its
variations across the universe.

To conclude we note that our results are consistent with previous work
based on completely different techniques and samples when reframed in
terms of a global IMF \citep[as summarized by,
e.g.,][]{Treu2010,Cap+12}. Furthermore our hypothesis of a different
IMF for bulge and disk is consistent with the upper limits on total
mass within the Einstein radius for the entire SWELLS and SLACS
samples, presented in a companion paper \citep[][in
preparation]{Brewer2012b}. Whereas SWELLS-III showed that the SWELLS
and SLACS data are consistent with an IMF that changes as a function
of galaxy stellar mass of stellar velocity dispersion,
\citet{Brewer2012b} shows that a scenario where the IMF is
Salpeter-like in the bulge and Chabrier-like in the disk is perfectly
consistent with the lensing constraints.


\section*{Acknowledgements}
AAD acknowledges financial support from the National Science
Foundation Science and Technology Center CfAO, managed by UC Santa
Cruz under cooperative agreement No. AST-9876783. AAD was partially
supported by HST grants GO-12292, GO-11978, and GO-11202.
TT acknowledges support from the NSF through CAREER award NSF-0642621,
and from the Packard Foundation through a Packard Research Fellowship.
PJM was given support by the Royal 
Society in the form of a research fellowship.
MB acknowledges support from the Department of Energy contract
DE-AC02-76SF00515.
DCK acknowledges support from HST grant GO-11206.02-A, and NSF grant
AST-0808133.
LVEK acknowledges the support by an NWO-VIDI programme subsidy
(programme number 639.042.505).
This research is supported by NASA through Hubble Space Telescope
programs GO-10587, GO-10886, GO-10174, 10494, 10798, 11202, 11978,
12292 and in part by the National Science Foundation under Grant
No. PHY99-07949. and is based on observations made with the NASA/ESA
Hubble Space Telescope and obtained at the Space Telescope Science
Institute, which is operated by the Association of Universities for
Research in Astronomy, Inc., under NASA contract NAS 5-26555, and at
the W.M. Keck Observatory, which is operated as a scientific
partnership among the California Institute of Technology, the
University of California and the National Aeronautics and Space
Administration. The Observatory was made possible by the generous
financial support of the W.M. Keck Foundation. The authors wish to
recognize and acknowledge the very significant cultural role and
reverence that the summit of Mauna Kea has always had within the
indigenous Hawaiian community.  We are most fortunate to have the
opportunity to conduct observations from this mountain.
Funding for the SDSS and SDSS-II was provided by the Alfred P. Sloan
Foundation, the Participating Institutions, the National Science
Foundation, the U.S. Department of Energy, the National Aeronautics
and Space Administration, the Japanese Monbukagakusho, the Max Planck
Society, and the Higher Education Funding Council for England. The
SDSS was managed by the Astrophysical Research Consortium for the
Participating Institutions. The SDSS Web Site is http://www.sdss.org/.


\label{lastpage}
\bsp

\end{document}